\documentclass{agujournal}
\journalname{Space Weather}
\usepackage{latexsym}
\usepackage{graphicx}

\begin{document}

\title{Forecasting Periods of Strong Southward Magnetic Field Following Interplanetary Shocks}

\authors{T.\ M.~Salman\affil{1}\thanks{Department of Physics and Space Science Center, University of
New Hampshire, Morse Hall, 8 College Rd, Durham, NH 03824, USA.},
 N.\ Lugaz\affil{1}, C.\ J.~Farrugia\affil{1},
 R.\ M.~Winslow\affil{1}, A.\ B.~Galvin\affil{1}, and N.\ A.~Schwadron\affil{1}}

\affiliation{1}{Space Science Center and Department of Physics, University of New Hampshire, Durham, NH, USA.}

\correspondingauthor{T. M. Salman}{ts1090@wildcats.unh.edu}

\begin{keypoints}
\item Establish a probabilistic ensemble forecast for strong southward B$_{z}$ using fast-forward shocks, solar wind, and IMF parameters in the previous 24 hours.
\item Quantify the best combination of parameters to identify past analogues: B, B$_{z}$ and N$_{p}$ with most weight on the 30 minutes interval around the shock arrival.
\item Gives on average a 14-hour warning, has a skill score of 0.64 on average, and is an improvement over any forecast with set probability.
\end{keypoints}

\begin{abstract}

Long periods of strong southward magnetic fields are known to be the primary cause of intense geomagnetic storms. The majority of such events are caused by the passage over Earth of a magnetic ejecta. Irrespective of the interplanetary cause, fast-forward shocks often precede such strong southward B$_{z}$ periods. Here, we first look at all long periods of strong southward magnetic fields as well as fast-forward shocks measured by the \textit{Wind} spacecraft in a 22.4-year span. We find that 76{\%} of strong southward B$_{z}$ periods are preceded within 48 hours by at least a fast-forward shock but only about 23{\%} of all shocks are followed within 48 hours by strong southward B$_{z}$ periods. Then, we devise a threshold-based probabilistic forecasting method based on the shock properties and the pre-shock near-Earth solar wind plasma and interplanetary magnetic field characteristics adopting a ``superposed epoch analysis''-like approach. Our analysis shows that the solar wind conditions in the 30 minutes interval around the arrival of fast-forward shocks have a significant contribution to the prediction of long-duration southward B$_{z}$ periods. This probabilistic model may provide on average a 14-hour warning time for an intense and long-duration southward B$_{z}$ period. Evaluating the forecast capability of the model through a statistical and skill score-based approach reveals that it outperforms a coin-flipping forecast. By using the information provided by the arrival of a fast-forward shock at L1, this model represents a marked improvement over similar forecasting methods. We outline a number of future potential improvements.

\end{abstract}


\section{Introduction} \label{sec:intro}

Strong interaction of the solar wind with Earth's magnetosphere can give rise to geomagnetic storms. \citet{Fairfield:1966} identified the southward component of the interplanetary magnetic field (IMF) to be associated with ground magnetic disturbances on Earth. \citet{Echer:2008} studied 90 intense geomagnetic storms (when the disturbed storm time index, Dst reached below -100~nT) during solar cycle 23 (1996-2006) and found all of them to be associated with strong southward B$_{z}$. Southward B$_{z}$ opens the subsolar magnetopause through magnetic reconnection which allows the transfer of energy, plasma, and momentum to the magnetosphere \citep[]{Dungey:1961}. A criterion for intense geomagnetic storms is the presence of a long period (\textgreater 3h) of large, and negative (\textless -10~nT) IMF B$_{z}$ \citep[]{Gonzalez:1987}. While not infallible (see e.g., \citet{Farrugia:1998}), this is generally a very useful criterion. 

The driver of such storms may be a coronal mass ejection (CME), the sheath of shocked plasma upstream of a CME, corotating interaction regions (CIRs), or a combination of these structures \citep[]{Zhang:2007, Kilpua:2017}. CMEs are associated with large amounts of material ejected from the solar atmosphere into the solar wind. In particular, CMEs, and their subset magnetic clouds (MCs) are amongst the drivers of the strongest geomagnetic storms. Magnetic clouds are regions of enhanced magnetic field strength, smooth rotation of the magnetic field vector, and low proton temperature \citep[]{Burlaga:1981}. The reasons why CMEs drive the strongest geomagnetic storms are: the southward component of magnetic field in CMEs is non-fluctuating, can be especially strong \citep{Huttunen:2005,Zhang:2007}, and can last for several hours \citep[]{Farrugia:1997}. \citet{Zhang:2007} investigated 88 intense geomagnetic storms (Dst$\leq$-100 nT) that occurred during 1996-2005 and found the majority of them (87{\%}) to be caused by either single CMEs or multiple CMEs. \citet {Ontiveros:2010} studied 47 geomagnetic storms during the ascending phase of solar cycle 23 and found all of them to be associated with the passage of a shock followed by a CME. The orientation of IMF B$_{z}$ along with the impulsive energy at the CME front (or CME shock) determines the severity of space weather or intensity of geomagnetic storms \citep[]{Balan:2014}.

On the other hand, CIRs are consequences of spatial variability in the coronal expansion and solar rotation, which radially aligns solar wind flows of different speeds \citep[]{Gosling:1999}. CIRs correspond to weak or moderate storms due to their fluctuating magnetic and velocity fields in the overtaking high-speed stream, consistent with Alfvenic waves \citep[]{Tsurutani:1995}. Either the CIR or the high-speed stream or both can be drivers of storms \citep[]{Borovsky:2006}.

Apart from the structures described above, shocks or sheaths can also be geoeffective \citep[]{Huttunen:2002}. Shocks can accelerate solar energetic particles \citep[]{Reames:1999} and the sheath regions downstream of the shocks containing IMF compressed by the shock can also drive intense geomagnetic storms \citep[]{Tsurutani:1988}. CME sheaths are responsible for a quarter \citep[]{Richardson:2001} to a half \citep[]{Tsurutani:1988} of all geomagnetic storms. \citet{Lugaz:2016} highlighted multiple ways for a shock-sheath structure to be geoeffective.

Intense geomagnetic storms are often preceded by storm sudden commencements (SSCs). SSCs serve as an indicator of the first impact of a travelling disturbance associated with a strong dynamic pressure change on the magnetosphere \citep[]{Joselyn:1990}. The disturbance compresses the magnetosphere resulting in an abrupt increase of the Dst \citep[]{Ontiveros:2010}. What causes SSCs is mainly the shock waves driven by fast CMEs \citep[]{Russell:1974, Echer:2004, Echer:2008}. CIR-driven storms generally lack sudden commencements \citep[]{Kamide:1998}. 

Strong shocks are usually found ahead of fast CMEs \citep[]{Borovsky:2006}. Thereby most SSCs are associated with strong interplanetary shocks \citep[]{Iucci:1988, Russell:1992}. \citet{Gold:1955} initially suggested the association of SSCs with interplanetary shocks, which was validated by later  studies. \citet{Chao:1974} showed that out of 48 SSCs where interplanetary data are available at solar maximum between 1968 and 1971, 41 were associated with shock events. In a similar study, \citet{Smith:1986} found nearly a one-to-one correspondence between SSCs and interplanetary shocks over solar maximum between 1978 and 1981. \citet{Gosling:1991} found all but one of the 37 largest geomagnetic storms in the period of 1978 to 1982 to be associated with the passage of a CME and/or shock disturbances at Earth and the large majority of them (27 out of 37) to be associated with interplanetary events where Earth encountered both a shock and the CME driving it. However, CMEs/shocks are not a sufficient cause behind intense geomagnetic storms because it has also been shown that half of all CMEs and half of all shock disturbances encountered by Earth in this span did not produce any substantial geomagnetic activity. 

Due to the existence of a strong correlation between southward B$_{z}$ periods and geomagnetic storms, reliable prediction of B$_{z}$ can be thought of as a primary basis for accurate space weather forecasting. Several studies have been carried out to predict B$_{z}$ from purely statistical approaches to physics-based models and all manner of hybrids in between based on coronal data or near-Earth solar wind observations or using observations of B$_{z}$ in conjunction with other solar wind parameters to form so-called ``coupling functions''. However, as the IMF frequently fluctuates, predicting southward B$_{z}$ is full of uncertainties \citep[]{Jurac:2002}. \citet{Chen:1997} argued that for particularly well-defined magnetic structures, B$_{z}$ could be predicted up to approximately 10 hours in advance. These structures such as magnetic clouds have a smooth rotation of the IMF observed at L1. This allows for the extrapolation of B$_{z}$ to later times when the event reaches Earth, but such a forecasting scheme has not been tested in a quantitative way yet. \citet{Kim:2014} demonstrated a two-step forecasting technique to improve the forecast capability for geomagnetic storms. Their specific framework is based on estimating initial CME parameters and updating the forecast through monitoring of near real-time solar wind conditions. It sets out an interesting framework for considering how to best approach event-based prediction and uses a combination of parameters and approaches to address shortcomings of each other. \citet{Savani:2015} used a combination of remote sensing and empirical relationships for predicting magnetic vectors of magnetic clouds when they reach Earth. This approach builds on the classification scheme of \citet{Bothmer:1998}, which itself uses the helicity rule of \citet{Rust:1994}. Related but different techniques have been proposed by \citet{Palmerio:2017, Moestl:2018}, which rely purely on solar and coronal observations to determine the initial ``state'' of the CME. However, these techniques may not be always successful in forecasting complex geomagnetic storms (for example due to multiple CMEs) and need to be further tested over a large number of events. \citet{Balan:2017} suggested a scheme for forecasting severe space weather based on the sudden increase in solar wind velocity due to a high-velocity CME front (or CME shock) coinciding with large southward B$_{z}$. This forecasting scheme provides a forecast time of less than 35 minutes. \citet{Riley:2017} outlined a pattern recognition (PR) technique for forecasting B$_{z}$. Using a mean square error method, they found that the PR technique shows only limited improvement over the baseline model (B$_{z}$=0 always) for the prediction of B$_{z}$. They also observed that the prediction correlation generally increases for a small forecasting window of 6 hours. \citet{Owens:2017} demonstrated an analogue ensemble (AnEn) method for probabilistic forecasting of B$_{z}$. They used a cost/loss method for determining the effectiveness of a probabilistic forecast. They found that AnEn method for probabilistic forecasting of B$_{z}$ has value only for relatively small forecast lead times (3 hours) and for large negative values of B$_{z}$ (\textless -2.93~nT).
 
\citet{Jurac:2002} examined the influence of shock parameters on geomagnetic disturbances within 48 hours following the shock arrival at Earth. They studied 107 fast-forward shocks observed by the \textit{Wind} spacecraft from 1995 to 2000. They identified the angle between the shock normal and the upstream IMF, $\theta_{u}$ to be a good indicator of future geoeffectivness of a shock. They found that shocks with $\theta_{u}$ between 70--90$^{\circ}$ are followed by intense storms 38--50{\%} of the time. They found an even better correlation using the angle between shock normal and the downstream IMF, $\theta_{d}$. They found that shocks with front normals orthogonal to the downstream IMF ($\theta_{d}$\textgreater 80$^{\circ}$, i.e., perpendicular shocks) are likely to be followed by intense storms 40{\%} of the time. This study highlighted the potential for some fast-forward shocks to provide an ``advanced warning'' of an incoming strong southward B$_{z}$ period depending on the shock parameters. To the best of our knowledge, this study has not been followed up to include more shocks and it is not currently used for real-time space weather forecasting.   

In this paper, we forecast strong southward B$_{z}$ periods that are preceded by fast-forward shocks. To do so, we construct a forecasting model by considering an ensemble of past analogues, similar to the procedure of \citet{Owens:2017}. This builds upon the assumption that past variability can serve as an indicator of future variability \citep[]{Riley:2017}. Our initial hypothesis is that the majority of strong southward B$_{z}$ periods are caused by CMEs, as found by previous studies, and we decide to build our ensemble of analogues only from periods around fast-forward shocks. It is known that CME speed is somewhat correlated with CME magnetic field strength \citep[]{Moestl:2014}. Slow CMEs and minor shock disturbances are not associated with significant geomagnetic activity \citep[]{Gosling:1991}. Thereby, fast CMEs can be expected to drive the majority of the strong southward B$_{z}$ periods. Near 1 AU, CMEs are the main drivers of fast-forward shocks \citep[]{Berdichevsky:2000}. These periods are therefore likely to be preceded by a fast-forward shock, driven by the CME. The level of geomagnetic activity stimulated by Earth passage of a shock disturbance or CME is related directly to the magnitude of the flow speed, magnetic field strength, and southward field component associated with the event \citep[]{Russell:1973, Akasofu:1981, Baker:1984}. As a result, prediction of these variable solar wind conditions in near-Earth space (i.e., at 1 AU) is important for space weather forecasting \citep[]{Owens:2005}. 

The paper is organized as follows. Data and methodology of our study are described in Section~\ref{sec:methodology}. We present the probabilistic forecast model for two reference events in Section~\ref{sec:example}. In Section~\ref{sec:results}, we find the best method for the prediction of strong southward B$_{z}$ periods. Then, we examine the probabilistic forecasts using a statistical and skill score-based approach. A brief summary and discussion are stated in Section~\ref{sec:conclusion}.

\section{Methodology} \label{sec:methodology}

Our model uses plasma and interplanetary magnetic field measurements for the solar wind around the arrival of a fast-forward shock at L1 (see Section~\ref{ssec:data} for details). We use the association between fast-forward shocks and strong and long-duration southward B$_{z}$ periods as the basis of our forecasting model. As such, we are only able to forecast the arrival of a long-duration (more than 3 hours) strong southward B$_{z}$ (B$_{z}$\textless -10~nT) following fast-forward shocks. Section~\ref{ssec:proportion} provides a quantitative justification of limiting the model to periods following fast-forward shocks only.

At the core of our technique is to find closest past events to the event that we wish to forecast. To do so, we select the interval of 24 hours prior to the shock to 0.25 hours after the shock arrival as the training window and the interval of 0.25 hours to 48 hours after the shock arrival as the forecast window as explained in Section~\ref{ssec:match}. For any event to be forecasted, we use the training window to determine the closest matches to the current solar wind conditions and use the observations of the post-shock intervals of these closest matches to make a probabilistic forecast for the reference event. These closest matches are found through quantifying variations of four solar wind and interplanetary magnetic field parameters in the training window using the root mean square error (RMSE) approach.

Once the closest matches are determined, we calculate how many of these reached our southward B$_{z}$ threshold (B$_{z}$\textless -10~nT for three consecutive hours or more) after the shock. The proportion of these closest matches that reach the threshold over the total number of closest matches determines the probabilistic forecast (see Section~\ref{ssec:error} for details). 

We adopt some modifications to our technique of finding closest past events. We split the training window into two intervals as described in Section~\ref{ssec:modifications}. We assign different weights to the two intervals in the training window (Section~\ref{sssec:twwcp}). RMSE of the four parameters in both intervals are multiplied with weighting constants (Section~\ref{sssec:weights}).

In Section~\ref{ssec:threshold}, we convert the probabilistic forecasts into dichotomous forecasts by imposing a threshold criterion onto them. 

\subsection{Data} \label{ssec:data}

For our study, we select the time period of more than 22 years from 01/01/1995 to 05/27/2017. We use the shock database of \citet{Kilpua:2015}, generated and maintained at the University of Helsinki which can be found at: http://ipshocks.fi/ and the shock database of Harvard Smithsonian Center for Astrophysics which can be found at: https://www.cfa.harvard.edu/shocks/. Hereafter, we will refer to these two databases as IPSDB and CFASDB respectively for brevity. For all the fast-forward \textit{Wind} shocks with the fast magnetosonic Mach number, M$_{ms}$\textgreater 1 in this span, we select the 3-day window around a shock, containing the period from 24 hours prior to the shock to 48 hours after the shock arrival at L1. To record the solar wind conditions during these windows, we use 1-minute high resolution data from the \textit{Wind} Magnetic Field Investigation \citep[MFI,][]{Lepping:1995} instrument {\&} 90-second resolution plasma data from the Solar Wind Experiment \citep[SWE,][]{Ogilvie:1995} instrument. As explained below, we focus on four parameters: magnetic field magnitude (B), z-component of the magnetic field (B$_{z}$) in GSM (Geocentric Solar Magnetospheric) coordinates, solar wind proton number density (N$_{p}$), and x-component of the solar wind velocity (V$_{x}$) in GSM coordinates. Hereafter, we refer to these 3-day windows as ``events''. All of our events have the same time scale with $t=0$~hours corresponding to the shock arrival.

\subsection{Examining the Proportion of Strong Southward B$_{z}$ Periods Following Fast-Forward Shocks} \label{ssec:proportion}

We first examine the association between fast-forward shocks and subsequent strong and long-duration southward B$_{z}$ periods to determine if the proposed approach of focusing on periods following fast-forward shocks is appropriate. For this study only, we use hourly near-Earth OMNI data \citep[]{King:2005} of GSM B$_{z}$ component extracted from NASA/GSFC's OMNI data set through OMNIWeb.

All instances in the period 01/01/1995-05/27/2017 where B$_{z}$\textless -10~nT for 3 consecutive hours or more are listed. We find 129 such instances. Then we use the two shock databases to match these instances with fast-forward shocks (with M$_{ms}$\textgreater 1) observed by the \textit{Wind} spacecraft within the pre-48h interval of an instance. However, we mainly focus on IPSDB. Using IPSDB, we find that 81 (63{\%}) instances are preceded by at least one fast-forward shock occurring within the pre-48h interval. Then we use CFASDB to look for \textit{Wind} shocks which are not listed in IPSDB but may be preceding the 48 negative (not preceded by any fast-forward shock occurring within the pre-48h interval) instances. We find that 6 of these negative instances are indeed preceded by a fast-forward shock not listed in IPSDB but in CFASDB. We then use both IPSDB and CFASDB to look for ACE shocks which may have preceded the 42 remaining negative instances. We find that 11 of these are preceded by a fast-forward shock observed by the ACE spacecraft at L1. In total we find that, 98 of these instances (76{\%}) are preceded by at least one fast-forward shock occurring within the pre-48h interval. This validates the assumption that a large portion of the strong and long-duration southward B$_{z}$ periods are preceded by fast-forward shocks. We also have 2 negative instances on 10/29/2013 and on 10/30/2013 which have been reported as being preceded by shocks but considerable data gaps in the pre-48h interval might have lead to a no shock observation.      

We now look at the reverse association in the period 01/01/1995-05/27/2017. We treat multiple southward B$_{z}$ periods within a 24-h time interval as one. If multiple southward B$_{z}$ periods are preceded by the same shock, we count that shock only once. Also, there are instances where a single southward B$_{z}$ period is preceded by multiple shocks. There are 490 fast-forward shocks (with M$_{ms}$\textgreater 1 and listed in the IPSDB) observed by \textit{Wind} at L1 and we match them with 129 strong and long-duration southward B$_{z}$ periods. Out of these shocks, 100 are followed by strong and long-duration southward B$_{z}$ periods. Integration of the CFASDB and inclusion of ACE shocks from the IPSDB gives us 17 more such shocks (6 \textit{Wind} shocks from CFASDB and 11 ACE shocks from IPSDB). So, out of the 507 shocks studied in this time span, 117 (regardless of spacecraft type) are followed by strong and long-duration southward B$_{z}$ periods which accounts for 23{\%} of total. This leads to the understanding that, although the large portion of the strong and long-duration southward B$_{z}$ periods occur within 48 hours after the arrival of fast-forward shocks at L1, only a small portion of fast-forward shocks are actually followed by strong southward B$_{z}$ periods. We combine our shocks list with the ICME list of \citet{Richardson:2010} as done in \citet{Lugaz:2017a} and find that 86 ($\sim$74{\%}) of these 117 shocks are CME-driven shocks. So the majority of shocks followed by southward B$_{z}$ periods are driven by CME although CME-driven shocks only represent about 72{\%} of all fast-forward shocks at 1 AU \citep[]{Kilpua:2015}.

The time interval of our study is roughly 22.4 years. The forecast model is constructed to function by monitoring 3-day windows (1 day training window and 2 days forecast window) around a fast-forward shock arrival at L1. Removing the fast-forward shock concept from the picture, the forecast model has an uphill task of distinguishing 129 strong southward B$_{z}$ windows (3.2{\%} of total) from 4088 windows. This forecasting scheme was implemented by \citet{Riley:2017} with only moderate success to predict the southward magnetic field. In addition, testing different approaches based on the 22.4-year span of data, and the full range of solar wind plasma and IMF parameters with different weights would require a dedicated and complex data analysis study which may not improve significantly over their model. In our current approach as proposed here, we search for physics-based constraints to narrow down the total number of windows that are ``candidates'' as having strong southward B$_{z}$. We do so by considering those windows preceded by a fast-forward shock. Limiting ourselves to windows around fast-forward shocks enables to develop a model aimed at distinguishing the 23{\%} positive windows out of 507 instead of 3.2{\%} out of 4088. However, it is achieved at the cost of not being able to predict 24{\%} of strong southward B$_{z}$ events, those not preceded by a fast-forward shock.

We also expand upon the study of \citet{Jurac:2002} by considering the time period 1995--2017 rather than 1995--2000 and we carry out a similar examination of 507 fast-forward shocks during this period. We find 190 shocks with upstream angles between 70--90$^{\circ}$. However, only 42 of them are followed by an intense southward B$_{z}$ period within 48 hours after the shock arrival which accounts for 22{\%} of the total. For another shock parameter, the downstream angle, we find 200 shocks with downstream angles greater than 80$^{\circ}$ and 50 of them are followed by an intense southward B$_{z}$ period within 48 hours after the shock arrival which accounts for 25{\%} of the total. Therefore, the correlation between shock angles and geoeffectiveness observed by \citet{Jurac:2002} does not hold true for an expanded interval and increased number of shocks. The proportion we observe for shocks with angles between 70--90$^{\circ}$ is almost identical to the proportion we found for all fast forward shocks (23{\%} of fast-forward shocks are followed by intense storms within 48 hours after the shock arrival at L1 regardless of any parameter criteria). As our probabilistic forecast model is based on periods following fast-forward shocks, this finding leads us to consider different combinations of upstream parameters.

\begin{figure*}[htbp!]
 \centering
\includegraphics[width=\linewidth]{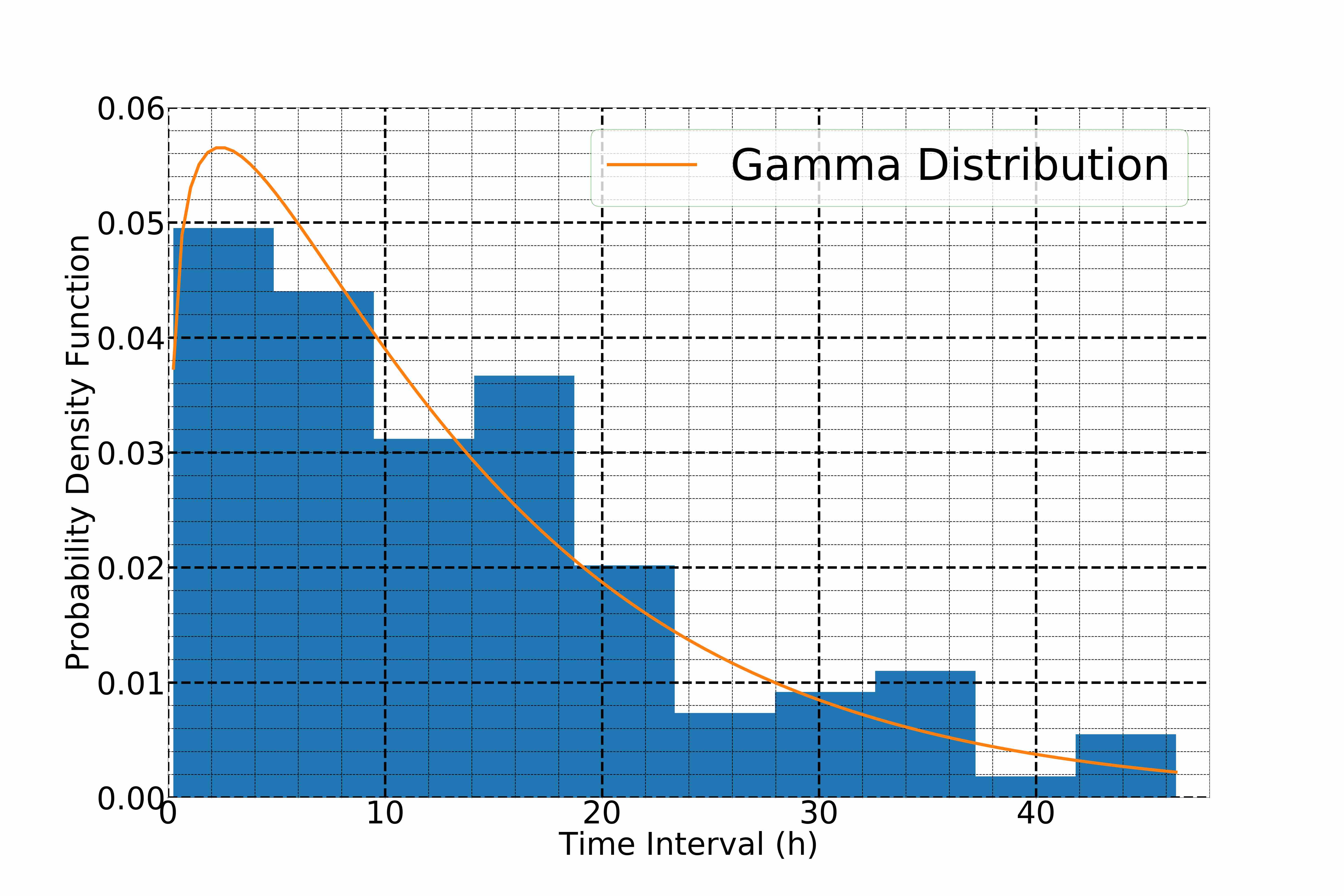}
  \caption{Distribution of time intervals between fast-forward shocks and the start time of their corresponding strong and long-duration southward B$_{z}$ periods.}
  \label{fig:distribution}
\end{figure*}

By examining the time intervals between fast-forward shocks and the start time of their corresponding strong and long-duration southward B$_{z}$ periods, we find the mean and median of these intervals to be 13.87 hours and 11.59 hours respectively. This indicates that a forecast of strong and long-duration B$_{z}$ periods based on shock arrival at L1 may give on average a 14-hour advanced warning. Examining these intervals more precisely, we observe a large range (from a minimum time interval of 15 minutes to a maximum time interval of approximately 47 hours). It is also seen that half of the time intervals lie in the range between 5.19 hours (25th percentile) and 19.15 hours (75th percentile). To first order this 25-75{\%} range corresponds to southward B$_{z}$ starting towards the back of a sheath or the beginning of the ejecta (typical sheath duration is 7.5 hours) to the second half of the ejecta (after a 7.5-hour sheath and 11.6 hours of the $\sim$20-hour magnetic ejecta). The shortest durations are for shocks inside CME, see \citet{Lugaz:2015}. The greater than 75{\%} may be at the back of a CME or in the wake or due to multiple CME interaction \citep[]{Lugaz:2017b}. Figure~\ref{fig:distribution} shows the theoretical distribution which best fits these time intervals. It is a Gamma distribution with shape parameter 1.67 and scale parameter 8.30.

\subsection{Identification of Closest Matches} \label{ssec:match}

At the core of the forecasting technique is the idea that the shock parameters may contain information related to the driver of the shock. For example, the speed of a CME-driven shock is certainly related to the speed of the CME front. As this method is meant to be used in near-real time, it is simpler and faster not to calculate the shock speed, normal direction, etc. from the Rankine-Hugoniot relationships but instead to rely on data from the shock ramp. As such, our ``training window'', where the closest matches to a reference event are identified, includes the period before the shock as well as the shock ramp itself. In details, we use the combination of the pre-shock (23.75 hours) interval and the shock itself ($\pm~0.25$ hours around the shock) as the training window and the post-shock (47.75 hours) interval as the forecast window. As we find from the time intervals that 90{\%} of strong and long-duration southward B$_{z}$ periods occur at least 1.85 hours after a shock arrival at L1, it enables us to include the post shock 0.25 hours period in the training window. For any reference event, we use the training window to determine the closest matches to the solar wind conditions to be forecasted and use observations of the corresponding post-shock intervals to make a probabilistic forecast for the reference event (see Section~\ref{ssec:error} for further details).

The closest matches to any reference event are found based on the variations of the four solar wind and IMF parameters of the database events in comparison with the reference event in the training window. These variations are quantified using the root mean square error (RMSE) method. We make the RMSE of a particular parameter dimensionless by dividing it by the average in the training window of that parameter of the reference event. All errors are done on dimensionless values because we do not want the RMSE of one of the parameters to dominate the calculation of TRMSE as the range of speed for example (250-1500 km/s) is very different than that of magnetic field (2-100 nT). The dimensionless RMSE for each of the four parameters (B, B$_{z}$, N$_{p}$, and V$_{x}$) are then added to get the total RMSE (TRMSE) given by:

\begin{equation}\label{eq:1}
TRMSE=\sum_{i=1}^{4} Z_{i}
\end{equation}

Z$_{i}$ is the dimensionless RMSE of a parameter, i runs from 1 to 4 (four parameters in consideration).

Based on the values of TRMSE (the lower the value of TRMSE, the closer is a event to the reference), the closest matches to a reference event are identified.

\subsection{Probabilistic Forecast and Error of Forecast} \label{ssec:error} 

With the closest matches to a reference event identified, we first make a probabilistic forecast for B$_{z}$ in the forecast window. Our focus is on B$_{z}$ reaching a pre-defined threshold, here -10~nT for 3 consecutive hours or more in the forecast window. The probabilistic forecast goes as follows:
If we focus on the N closest matches, and if M of those have B$_{z}$ reaching the threshold, a probability of M/N that the reference event may have B$_{z}$ reaching the threshold in its forecast window is assigned. For each probabilistic forecast, the error of the forecast is also calculated. If the reference event indeed reaches the threshold, the correct forecast is 1 and the probabilistic error is $\sqrt{1 - (M/N)^2}$. If the reference event does not reach the threshold, the correct forecast is 0 and the probabilistic error is M/N.
 
The primary goal of our probabilistic forecast model is to outperform the random (i.e., coin flipping) forecast which has an error of 0.5 irrespective of whether the event reaches the threshold or not. A perfect forecasting method would have an error of 0, error of 1 means no skill. We try to minimize the error of our probabilistic forecasts. As our statistics shows that 23{\%} of fast-forward shocks are precursors to a strong and long-duration B$_{z}$ period, the best naive guess is to assign a probability of 0.23 to every fast-forward shock for causing intense southward B$_{z}$. This best zero-skill approach leads to a RMS error of 0.421. Ideally, we would like to beat this, to show that our forecast model not only has more skill than a random forecast but also possesses more skill than the best possible average-based forecast.

\subsection{Modifications to the Probabilistic Forecast Model} \label{ssec:modifications}

In order to minimize the error, we randomly select 100 events from the 496 fast-forward shocks (490 listed in IPSDB + 6 additional from CFASDB) observed by \textit{Wind} in the period 01/01/1995-05/27/2017 and remove them from the database. This way the training set is fully different from the events set and cross-contamination is negated. We then randomly slice these 100 events into four individual sets (24, 25, 25 {\&} 26 events respectively). We use these events to form the basis of our model and refer to them as ``reference events'' from now on. The goal is to identify how to construct the TRMSE to minimize the error of forecasts. There is no mathematical or statistical justification for this unequal distribution of reference events. It is a mere coincidence. Among our first test set of 25 reference events, one event was removed due to a data gap issue. Then we conducted the analysis with the 24 reference events. Then expanding the results to four sets, we decided to add one more event in one of the other three sets.

\subsubsection{Training Window Weighting Constant Pair (TWWCP)} \label{sssec:twwcp}

\begin{figure*}[htbp!]
\centering
  \includegraphics[width=1.0\linewidth]{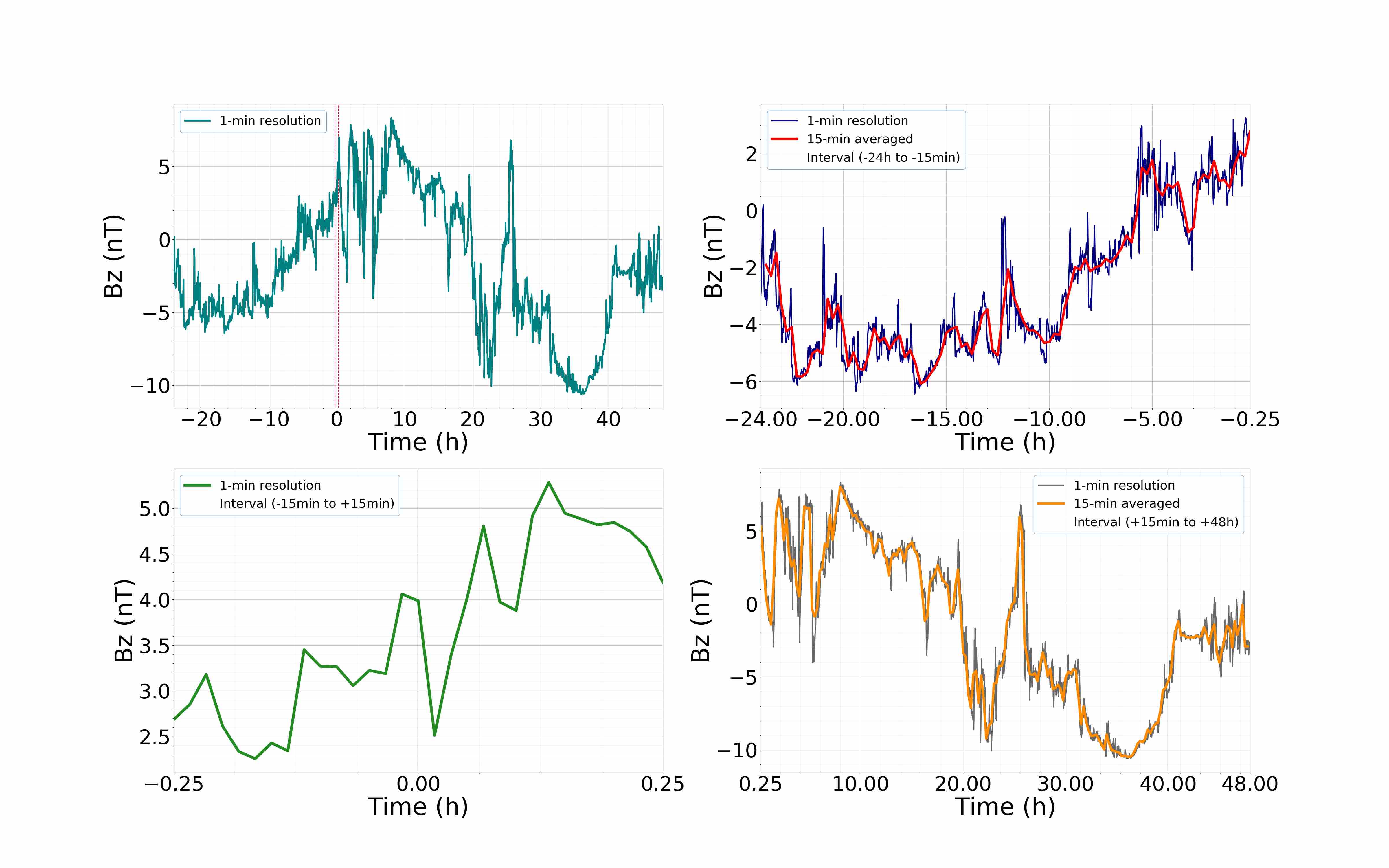}
  \caption{\textit{Wind} observations of the fast-forward shock on 25 May 1997. The panels show: i) variation of GSM B$_{z}$ component (top left) using 1-min resolution data over a 3-day window around the shock, from 24 May 1997 to 27 May 1997. The red dashed lines represent the interval of 0.25 hours prior to the shock arrival and 0.25 hours after the shock arrival ii) variation of GSM B$_{z}$ component (top right) using 1-min resolution data and 15-min averaged data for the interval of 24 hours to 0.25 hours prior to the shock arrival iii) variation of GSM B$_{z}$ component (bottom left) using 1-min resolution data for the interval of 0.25 hours prior to the shock arrival to 0.25 hours after the shock arrival iv) variation of GSM B$_{z}$ component (bottom right) using 1-min resolution data and 15-min averaged data for the interval of 0.25 hours to 48 hours after the shock arrival.}
  \label{fig:averaging}
\end{figure*}

The training window is split into two intervals: the first interval is the period of 24 hours to 0.25 hours prior to the shock and the second interval is the period of 0.25 hours prior to the shock to 0.25 hours after the shock arrival (it captures the shock jump). For the former interval, we bin the magnetic field and plasma data into 15-minute averages.  For the latter one, the 1-minute and 90-second data is used as it is. Figure~\ref{fig:averaging} shows the way the time interval is split and averaged.

For both intervals, we find the dimensionless RMSE of a parameter individually, multiply them with a weighting constant pair, and add them to get the RMSE for that particular parameter. Five sets of weighting constant pairs were tried: (1,0), (\( \frac{3}{4}\ ,\frac{1}{4}\)), (\( \frac{1}{2}\ ,\frac{1}{2}\)),(\( \frac{1}{3}\ ,\frac{2}{3}\)), and (\( \frac{1}{4}\ ,\frac{3}{4}\)).


A weighting constant pair ($m$,~1-$m$) means that in calculating the RMSE of a parameter in the training window, we are putting 100$m${\%} weight into the variations in the $-24$h to $-0.25$h interval and 100(1-$m$){\%} weight into the variations in the $-0.25$h to $+0.25$h interval. Introduction of the weigthing constant pair provides a key insight into the importance of both intervals to the prediction of southward B$_{z}$~periods. If the probabilistic forecast improves by adding more weight into the variations of a parameter in the $-0.25$h to $+0.25$h period in calculating the RMSE of that parameter in the training window, it suggests that what happens in the vicinity of the shock plays the prominent role in the occurrence of southward B$_{z}$~periods after the shock arrival. If the probabilistic forecast does not improve or worsens, it suggests the pre-shock interval is the important one. 

\subsubsection{Varying Weights of Parameter Variations} \label{sssec:weights}
 
Additionally, when the TRMSE is calculated by adding the dimensionless RMSE of the four solar wind and IMF parameters, different weights are assigned to each parameter. 
For example, a set of weights, $a_B$=1, $a_{Bz}$=1, $a_{Np}$=0, $a_{Vx}$=0, assigned to B, B$_{z}$, N$_{p}$, and V$_{x}$, respectively, indicates that the TRMSE in the training window is calculated with an equal weight for B and B$_{z}$, whereas the variations in N$_{p}$ and V$_{x}$ are neglected. These weights can have values ranging from 0 to 1 with 1 representing the maximum weight. We try out different sets of weights. In our model, the weight assigned to a particular parameter is same for both intervals of the training window. As such, the TRMSE is now given by:

\begin{equation}\label{eq:2}
TRMSE=\sum_{i=1}^{4} a_{i} [mZ_{1i} + (1-m)Z_{2i}]
\end{equation}

Z$_{1i}$ is the RMSE of a particular parameter in the pre-shock 23.75 hours period, Z$_{2i}$ is the RMSE of a particular parameter in the 0.5 hours period around the shock, and a$_{i}$ represents the weights assigned to the four parameters.

\subsection{Threshold Forecast in Conjunction with a Criterion} \label{ssec:threshold}

In addition to the probabilistic forecasts, it can be interesting to develop dichotomous forecasts (YES/NO) that assigns a probability of 100\% or 0\% to an event occurring. To do so, we impose a threshold onto our probabilistic forecasts. As our model provides probabilistic forecasts based on observations in the forecast windows of 15 closest matches, we have 15 thresholds to choose from. In our analysis, we extensively tried 5, 10, and 15 closest matches. Using the 15 closest matches to identify a threshold was found to work better than using the 5 or 10 closest matches. We did not try to determine if a number of closest matches larger than 15 would be better at forecasting strong B$_{z}$ period. As there are only 106 positive events in our database, many of them having different characteristics, using a larger number of closest matches is unlikely to improve the forecast much. We also found that using 10 or 15 closest matches was relatively similar. A more in-depth analysis of the most appropriate number of closest matches is left for a follow-up study.

Now, the thresholds in consideration can be generalized as \( \frac{n}{15} \) where n=1, 2, 3,...........15. Any probabilistic forecast exceeding that threshold will be considered a YES and any probabilistic forecast below that threshold will be considered a NO. Let us assume the threshold criterion selected for the probabilistic forecasts is 0.4 or \( \frac{6}{15} \). So any probabilistic forecast is considered a YES if it is greater than or equal to 0.4 (number of closest matches satisfying the B$_{z}$ criterion is $\geq$ 6) and is considered a NO if it is lower than 0.4 (number of closest matches satisfying the B$_{z}$ criterion is \textless~6).  

The skill of our threshold-based forecast model is quantified using the Heidke Skill Score (HSS, see Table~\ref{tab:0}) as reference \citep[]{Heidke:1926}. Negative skill score indicates that the model fairs worse than the random model, and positive skill score indicates the model is better than the random model. The perfect forecast will attain a HSS of 1 and a HSS of 0 means no skill.

\begin{table}[htbp!]
  \centering
  \caption{Contingency table for HSS.}
    \begin{tabular}{ccccc}
      \hline
          & Forecast & \multicolumn{2}{c}{Observed} &  \\\hline
          &       & YES   & NO    & Marginal Total \\\hline
          & YES   & a (\textit{Hit})     & b (\textit{False Alarm})     & a+b \\\hline
          & NO    & c (\textit{Miss})     & d (\textit{Correct Negative})    & c+d \\\hline
          & Marginal Total & a+c (\textit{Observed YES})     & b+d (\textit{Observed NO})     & a+b+c+d (\textit{Total}) \\\hline
          &  & \multicolumn{2}{c}{HSS = 2 (ad-bc) / [(a+c)(c+d) + (a+b)(b+d)] }       &\\\hline
    \end{tabular}%
  \label{tab:0}%
\end{table}%

\section{Example} \label{sec:example}

\begin{figure*}[htbp!]
 \centering
  \includegraphics[width=1.0\linewidth]{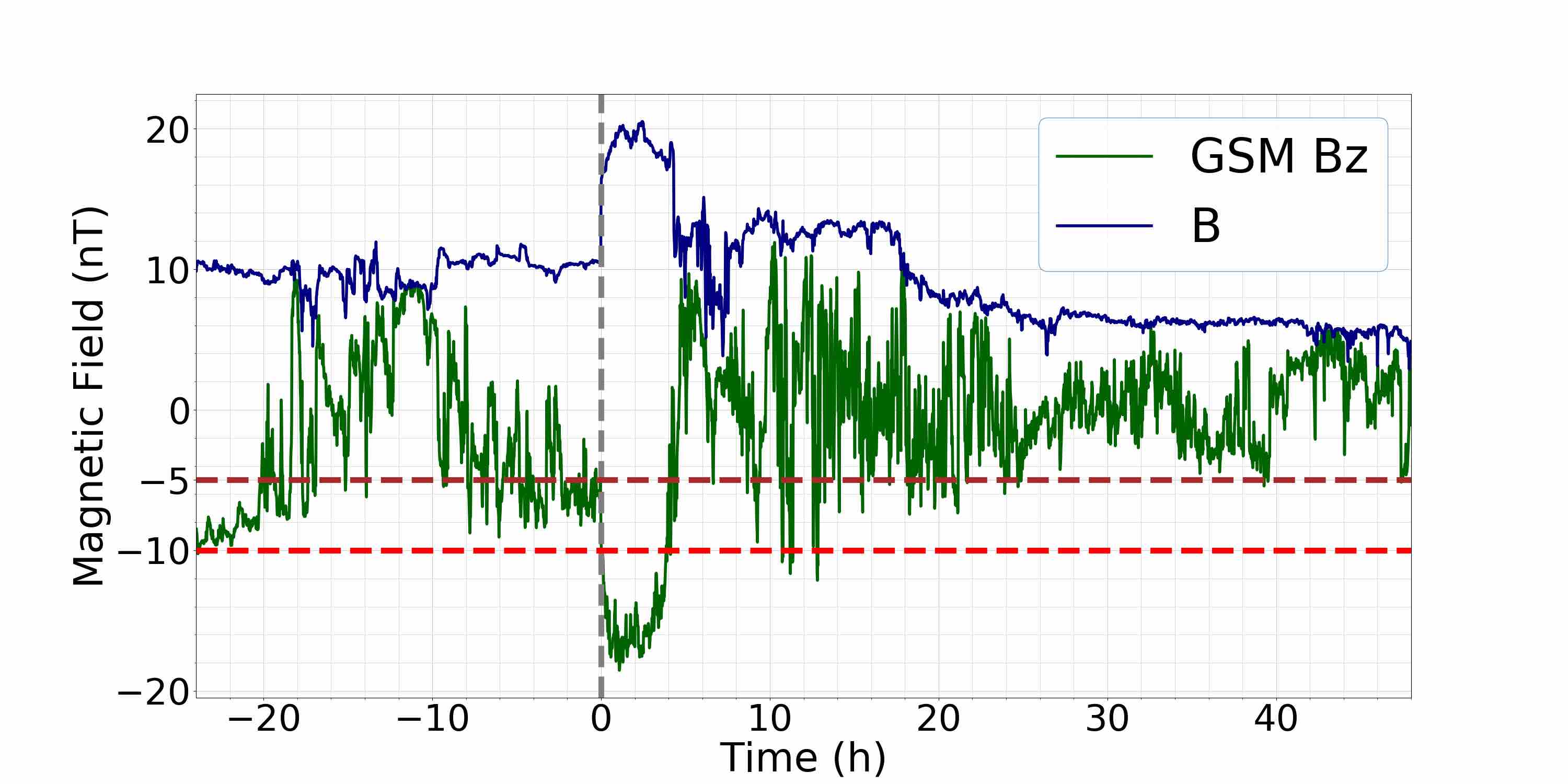}
  \caption{\textit{Wind} observations of the fast-forward shock on 18 June 2003. The panel shows the variation of magnetic field strength and GSM B$_{z}$ component over a 3-day window around the shock, from 17 June 2003 04:42 UT to 20 June 2003 04:42 UT.}
  \label{fig:reference event-I}
\end{figure*}

We demonstrate the forecasting model for a reference event observed by the \textit{Wind} spacecraft on 18 June 2003 at 04:42 UT (Figure~\ref{fig:reference event-I}). This reference event corresponds to a shock inside a CME \citep[]{Lugaz:2015, Lugaz:2016}. Though it is not a traditional event with a shock impacting Earth first followed by a magnetic ejecta, our motivation to present this event as an example is that it illustrates the pattern of improving probabilistic forecasts better than any of the other reference events we used in our analysis. Through this event, it is possible for us to showcase how including more than one parameter and putting more weight into the variations closer to the shock significantly improves the forecasts in a clear manner. 

For this reference event, the training window extends from 17 June 2003, 04:42 UT to 18 June 2003, 04:57 UT and the forecast window is the following 47.75 hours. Our goal is to predict the occurence of at least one B$_{z}$\textless -10~nT period of 3 consecutive hours or more within the forecast window. For this reference event, 17 minutes after the shock arrival, there is the start of a period of four consecutive hours where B$_{z}$\textless -10~nT. Our expectation from the probabilistic model is to filter out the 15 closest matches to the solar wind conditions of this reference event's training window and to generate a probabilistic forecast as close as possible to 1.
 
\begin{figure*}[htbp!]
  \centering
        \includegraphics[width=1.0\linewidth]{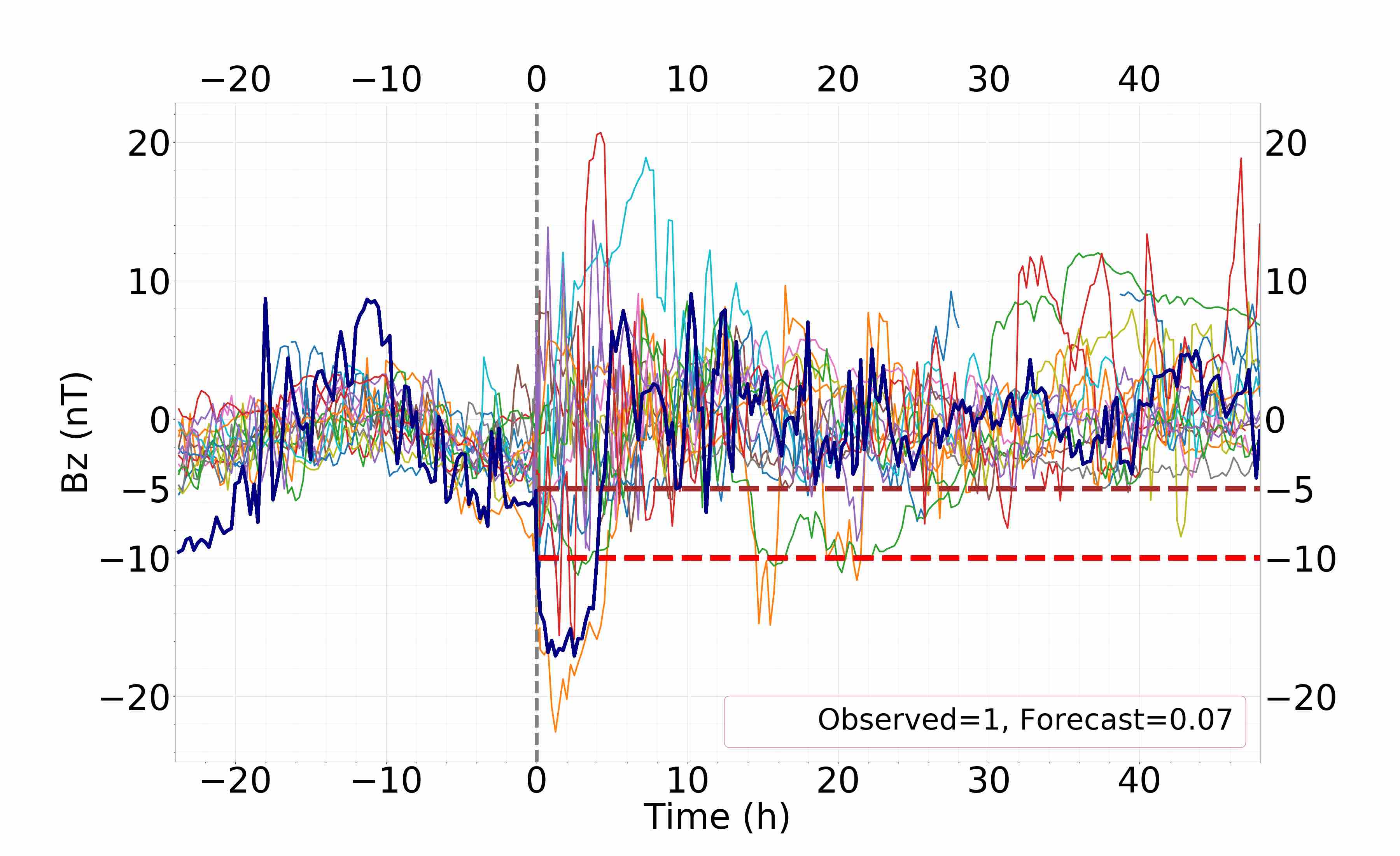}
        \caption{Probabilistic forecast for the event following the 18 June 2003 shock shown in Figure~\ref{fig:reference event-I} for TWWCP (1,0) and weights (0, 1, 0, 0) (Navy=reference event, other colors=15 closest matches).}
         \label{fig:forecast1}
  \end{figure*}

Initially, we start with our baseline (only considering B$_{z}$ variations and no shock info) set of weights: B$=$0, B$_{z}$=1, N$_{p}$=0, V$_{x}$=0 and choose (1,0) as the training window weighting constant pair (TWWCP). The TWWCP (1,0) is equivalent to (m,1-m) of Equation~\ref{eq:2}. It means that, when we calculate the TRMSE in the training window, we put 100{\%} weight into the variations in the $-24$h to $-15$min interval and 0{\%} weight into the variations in the $-15$min to $+$15min interval. The weights assigned to the four parameters are summarized in this order: $a_B$, $a_{Bz}$, $a_{Np}$, $a_{Vx}$ from now on. Choice of these weights and TWWCP suggests only variations of B$_{z}$  in the -24h to -15min interval prior to the shock arrival is considered. This set of weights and TWWCP gives a probabilistic forecast of 0.07 (Figure~\ref{fig:forecast1}).

\begin{figure*}[htbp!]
        \centering
        \includegraphics[width=1.0\linewidth]{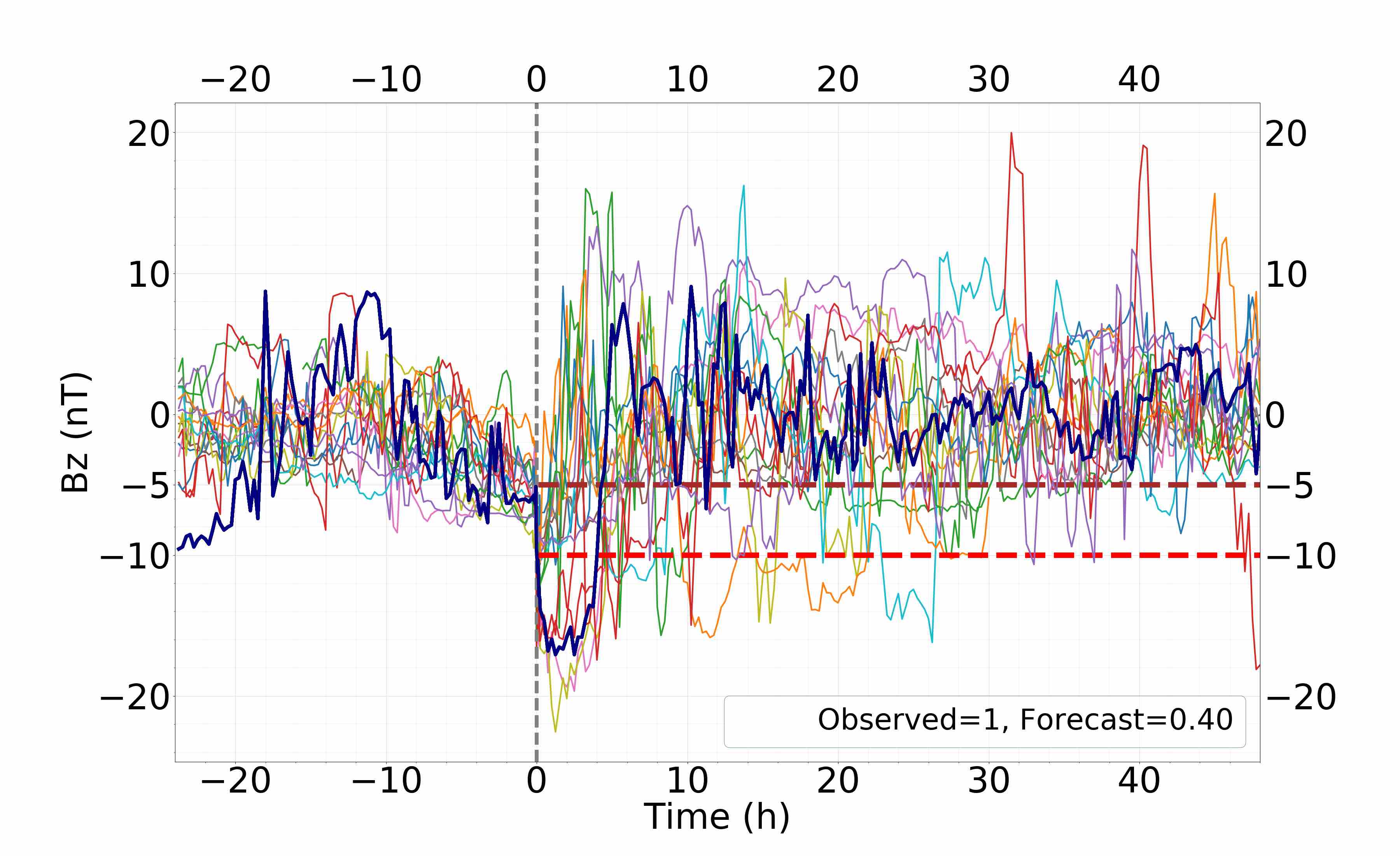}
        \caption{Same as Figure~\ref{fig:forecast1} but for TWWCP (\( \frac{1}{4}\ ,\frac{3}{4}\)).}
   \label{fig:forecast3}
\end{figure*}

This probabilistic forecast is far from ideal as the prediction is a 7{\%} probability for the occurence of B$_{z}$\textless -10~nT periods of 3 consecutive hours or more in the forecast window compared to an observed 100{\%} probability (as this reference event has a B$_{z}$\textless -10~nT period of 3 consecutive hours or more in its forecast window). To improve the probabilistic forecast, the TWWCP is changed from (1,0) to (\( \frac{1}{2}\ ,\frac{1}{2}\)) keeping the parameter weights fixed. It gives a probabilistic forecast of 0.33. This probabilistic forecast is an improvement for sure. Next, more weight is put into the variations of  B$_{z}$  closer to the shock by selecting TWWCP (\( \frac{1}{4}\ ,\frac{3}{4}\)) and using the same set of weights. It gives a probabilistic forecast of 0.40 (Figure~\ref{fig:forecast3}). This improvement of the probabilistic forecast from 0.07 to 0.33 and then to 0.4 shows that putting more weight into the variations closer to the shock improves the probabilistic forecast for this particular event.

\begin{figure*}[htbp!]
\centering
    \includegraphics[width=1.0\linewidth]{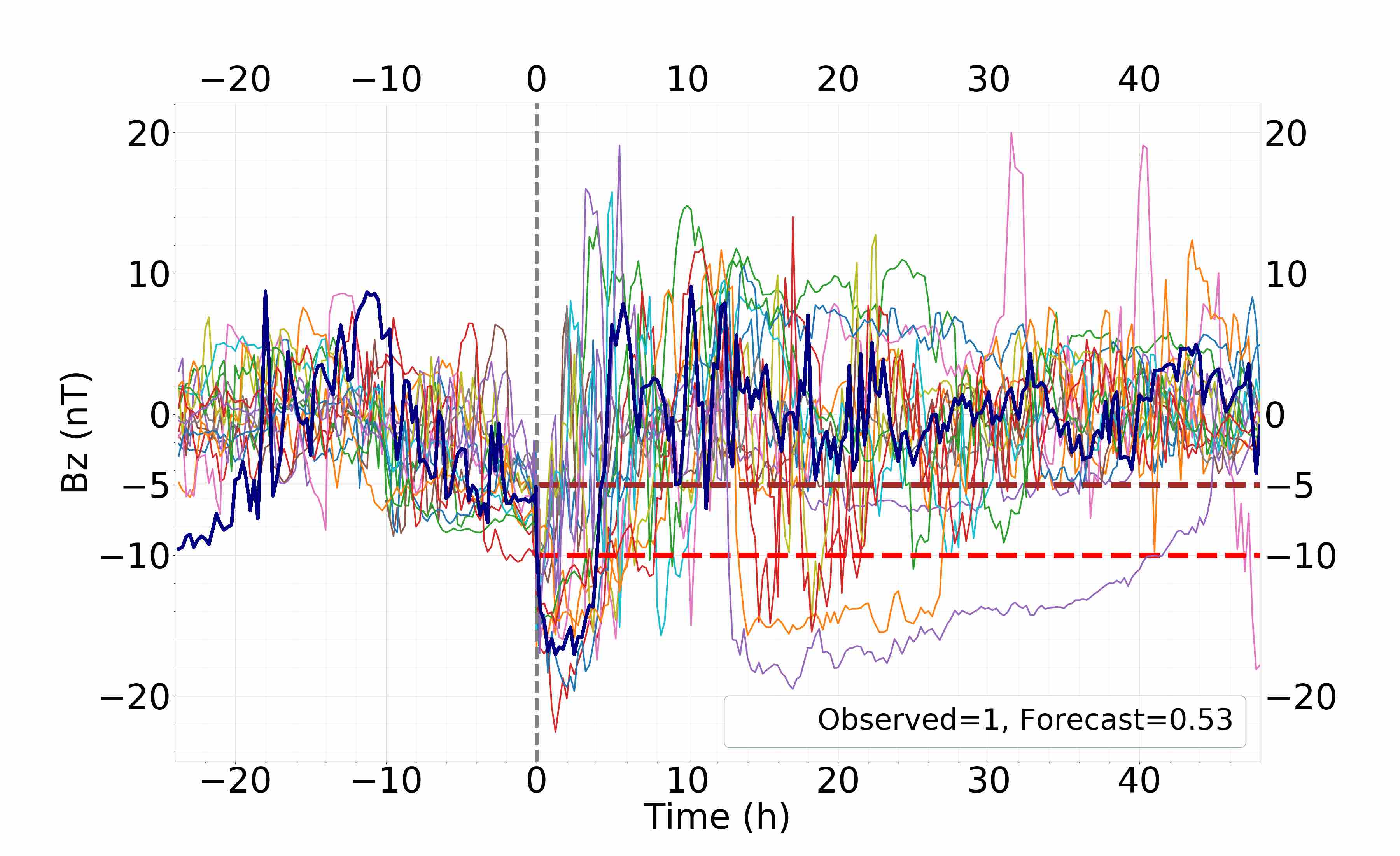}
    \caption{Same as Figure~\ref{fig:forecast1} but for TWWCP (\( \frac{1}{4}\ ,\frac{3}{4}\)) and weights (0.5, 1, 0, 0).}
     \label{fig:forecast6}
\end{figure*}

We monitor the evolution of probabilistic forecasts by changing the weights and TWWCPs. For each reference event, we accomplish this through numerous combinations of weights and TWWCPs. The goal of these repeated procedures is to identify the pattern of improving probabilistic forecasts. Thereby for this reference event, we now consider variations of both B and B$_{z}$ in the training window. Use of a set of weights (0.5, 1, 0, 0) and TWWCP (\( \frac{1}{4}\ ,\frac{3}{4}\)) gives a probabilistic forecast of 0.53 (Figure~\ref{fig:forecast6}).

This improvement in the probabilistic forecast indicates considering variations of more than one parameter in finding the 15 closest matches makes the probabilistic forecast better.

This assumption is validated by a probabilistic forecast of 0.13, previously 0.07 for the set of weights (0.25, 1, 0, 0) (previously (0, 1, 0, 0)) with TWWCP (1,0) and a probabilistic forecast of 0.47, previously 0.33 for the set of weights (1, 1, 0, 0) (previously (0, 1, 0, 0)) with TWWCP (\( \frac{1}{2}\ ,\frac{1}{2}\)). We also observe improvements in the probabilistic forecasts by increasing the weight of B.

\begin{figure*}[htbp!]
  \centering
    \includegraphics[width=1.0\linewidth]{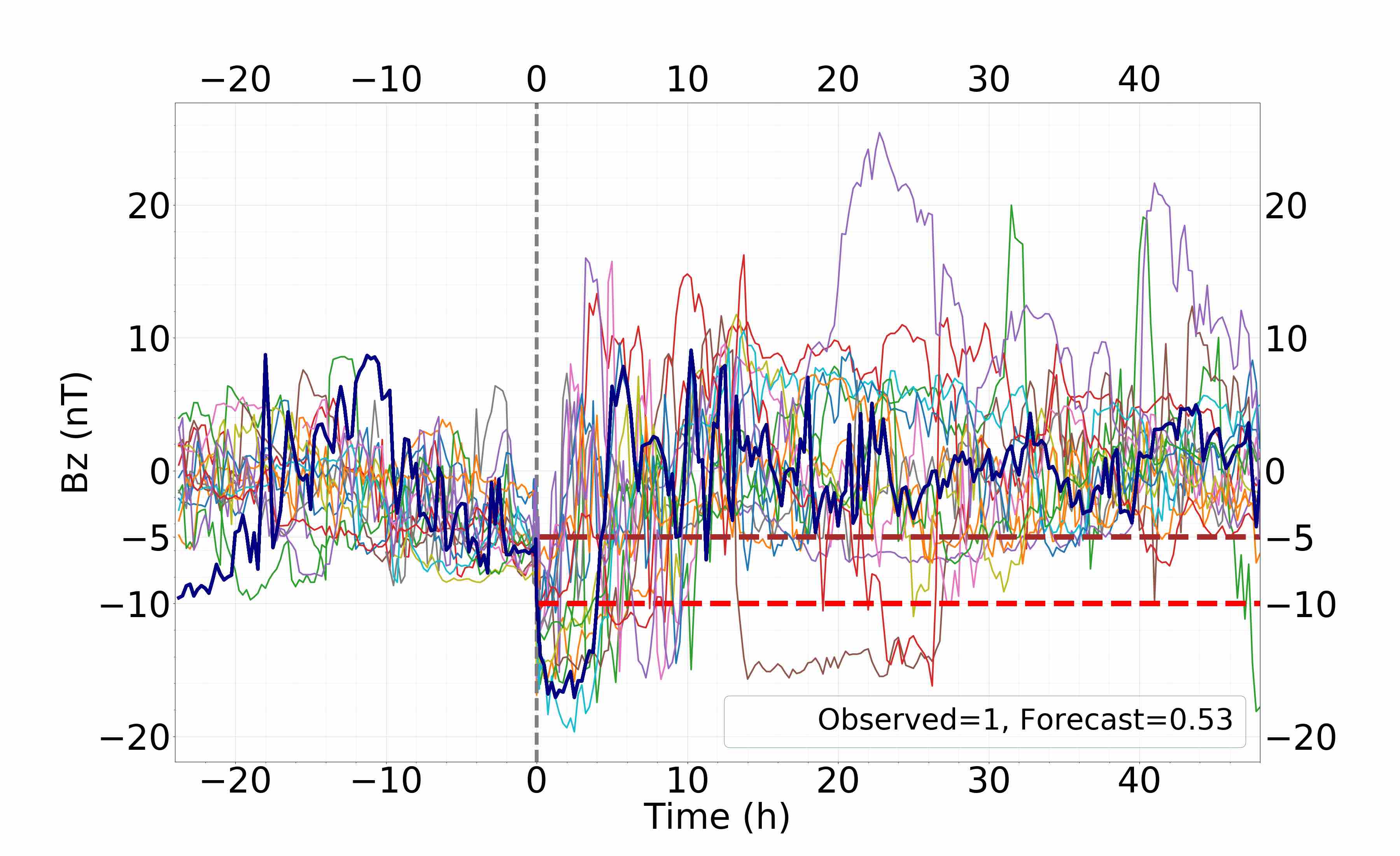}
    \caption{Same as Figure~\ref{fig:forecast1} but for TWWCP (\( \frac{1}{4}\ ,\frac{3}{4}\)) and weights (1, 1, 0.25, 0).}
     \label{fig:forecast7}
  \end{figure*} 

After that, variations of N$_{p}$ with the variations of B and B$_{z}$  in the training window is considered. The set of weights (1, 1, 0.25, 0) with TWWCP (\( \frac{1}{4}\ ,\frac{3}{4}\)) gives a probabilistic forecast of 0.53 (Figure~\ref{fig:forecast7}) and the set of weights (1, 1, 0.25, 0) with TWWCP (\( \frac{1}{2}\ ,\frac{1}{2}\)) gives a probabilistic forecast of 0.47. The general observed trend is that adding the parameter N$_{p}$ either keeps the forecast constant or improves it.

\begin{figure*}[htbp!]
  \centering
    \includegraphics[width=1.0\linewidth]{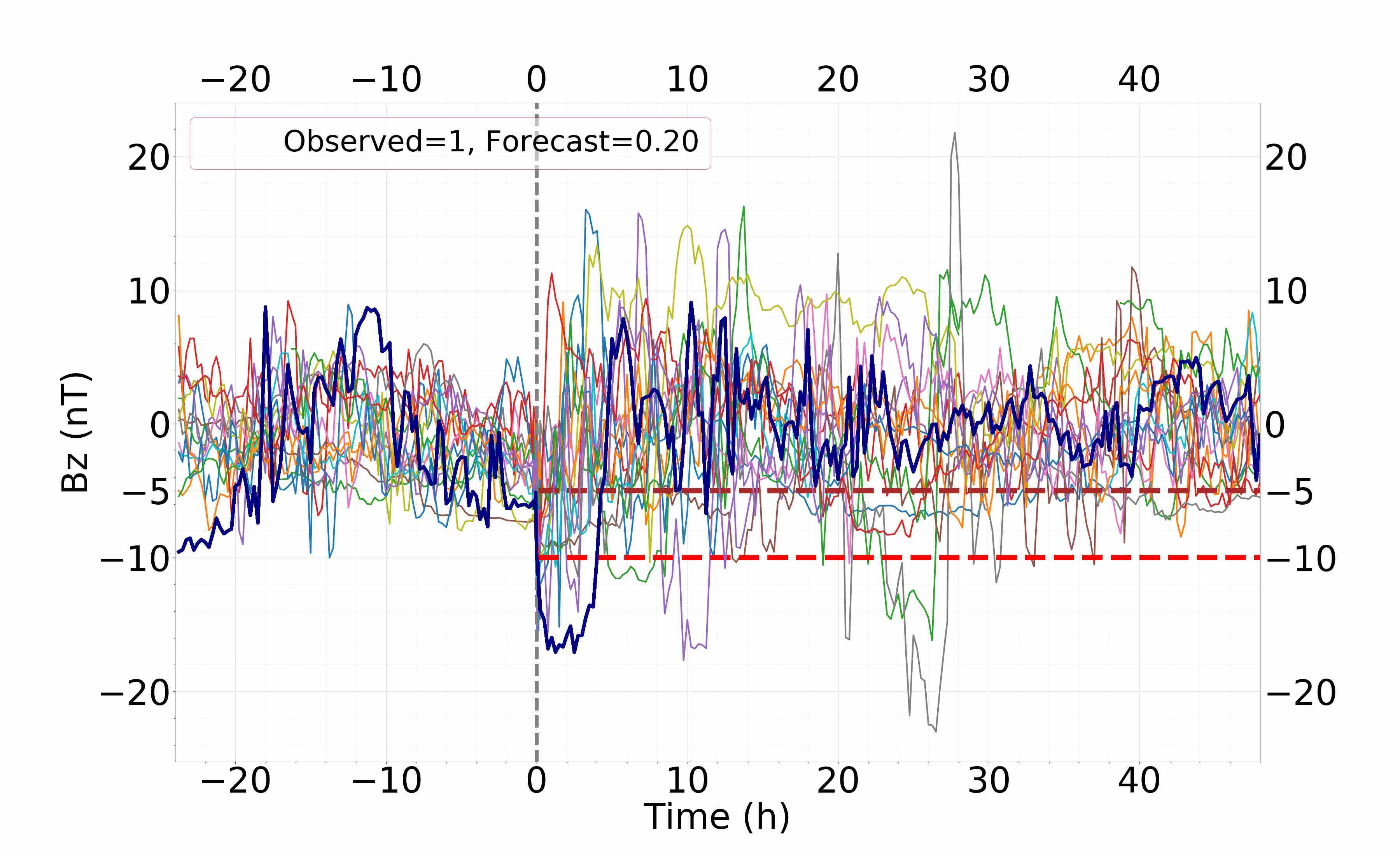}
    \caption{Same as Figure~\ref{fig:forecast1} but for TWWCP (\( \frac{1}{2}\ ,\frac{1}{2}\)) and weights (1, 1, 0, 0.5).}
     \label{fig:forecast8}
\end{figure*}

Finally, the influence of the fourth parameter (V$_{x}$) on the probabilistic forecasts is examined. The observed trend is that adding V$_{x}$ generally worsens the probabilistic forecasts. Two such scenarios are presented for a fixed TWWCP (\( \frac{1}{2}\ ,\frac{1}{2}\)). The probabilistic forecast of 0.47 for the set of weights (1, 1, 0, 0) drops to 0.2 (Figure~\ref{fig:forecast8}) for the set of weights (including V$_{x}$) (1, 1, 0, 0.5). Similarly a probabilistic forecast of 0.47 for the set of weights (1, 1, 0.25, 0) drops to 0.27 for the set of weights (including V$_{x}$) (1, 1, 0.25, 0.25). Important thing to note that this model only provides probabilistic predictions for the occurence of strong and long-duration southward B$_{z}$ periods following a fast-forward shock. The parameter V$_{x}$  may have influence on the prediction of  K$_{p}$ or Dst.

We also highlight the threshold-based forecast. As shown in Section~\ref{sec:results}, a good criterion is a probabilistic forecast of at least 0.4. 

For the reference event discussed above, the set of weights (1, 1, 0.25, 0) with TWWCP (\( \frac{1}{4}\ ,\frac{3}{4}\)) gives a probabilistic forecast of 0.53. So, for the threshold criterion of 0.40, this probabilistic forecast is considered a YES and it correctly predicts the occurence of B$_{z}$\textless -10~nT periods of 3 consecutive hours or more in the forecast window. Similarly, the set of weights (1, 1, 0, 0.5) with TWWCP (\( \frac{1}{2}\ ,\frac{1}{2}\)) gives a probabilistic forecast of 0.20. So for the threshold criterion of 0.40, this forecast is considered a NO and the outcome is a missed prediction.

\begin{figure*}[htbp!]
\centering
  \includegraphics[width=1.0\linewidth]{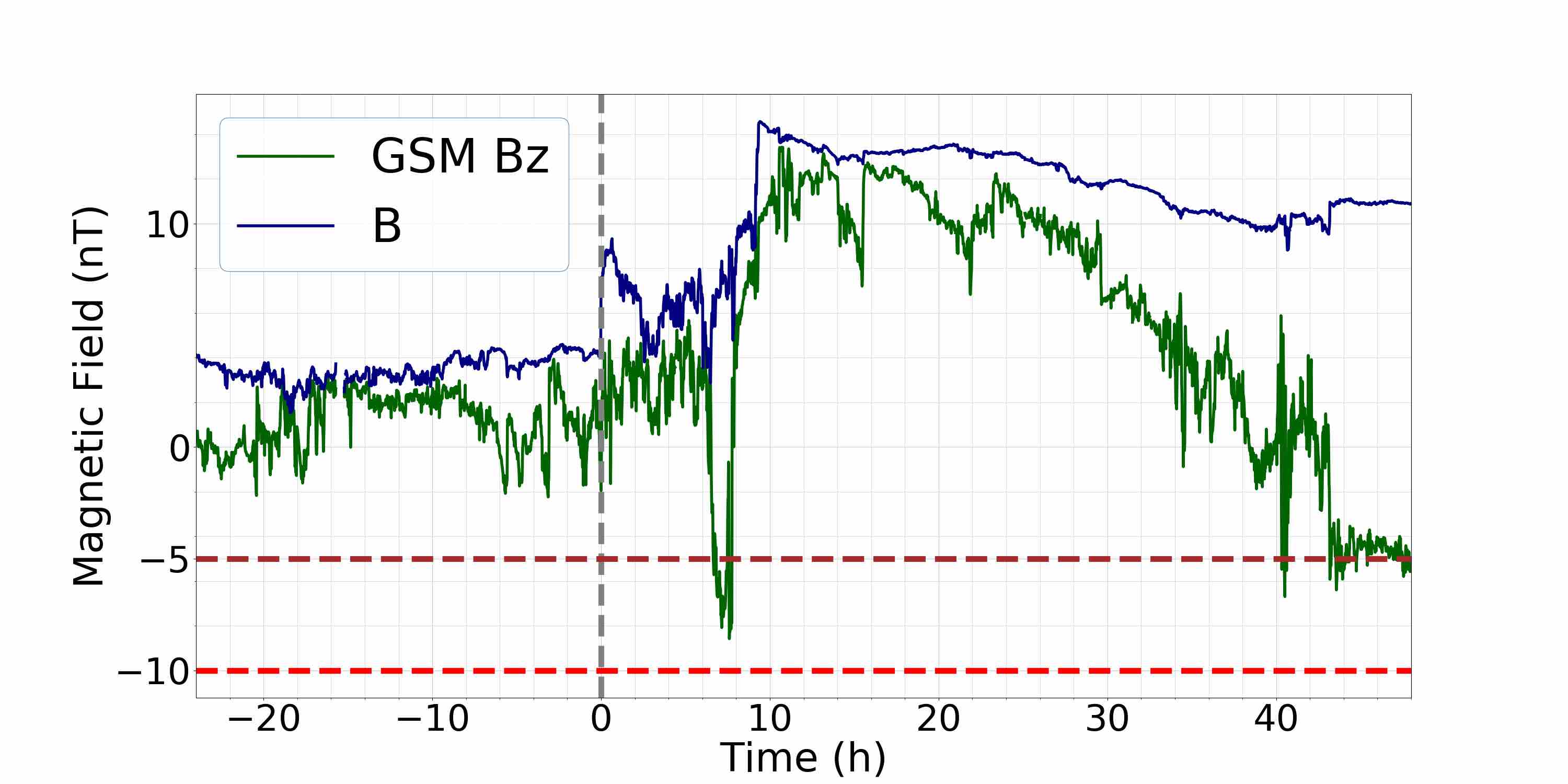}
  \caption{\textit{Wind} observations of the fast-forward shock on 29 March 2011. The panel shows the variation of magnetic field strength and GSM B$_{z}$ component over a 3-day window around the shock, from 28 March 2011 15:09 UT to 31 March 2011 15:09 UT.}
  \label{fig:reference event-II}
\end{figure*}

Figure~\ref{fig:reference event-II} represents a second reference event for which the model is demonstrated. This event was observed by the \textit{Wind} spacecraft on 29 March 2011 at 15:09 UT. This reference event lies in the opposite side of the spectrum compared to the first reference event as there is no B$_{z}$\textless -10~nT period of 3 consecutive hours or more in the forecast window. Thereby, the ideal probabilistic forecast needs to be as close to 0 as possible.

\begin{figure*}[htbp!]
  \centering
    \includegraphics[width=1.0\linewidth]{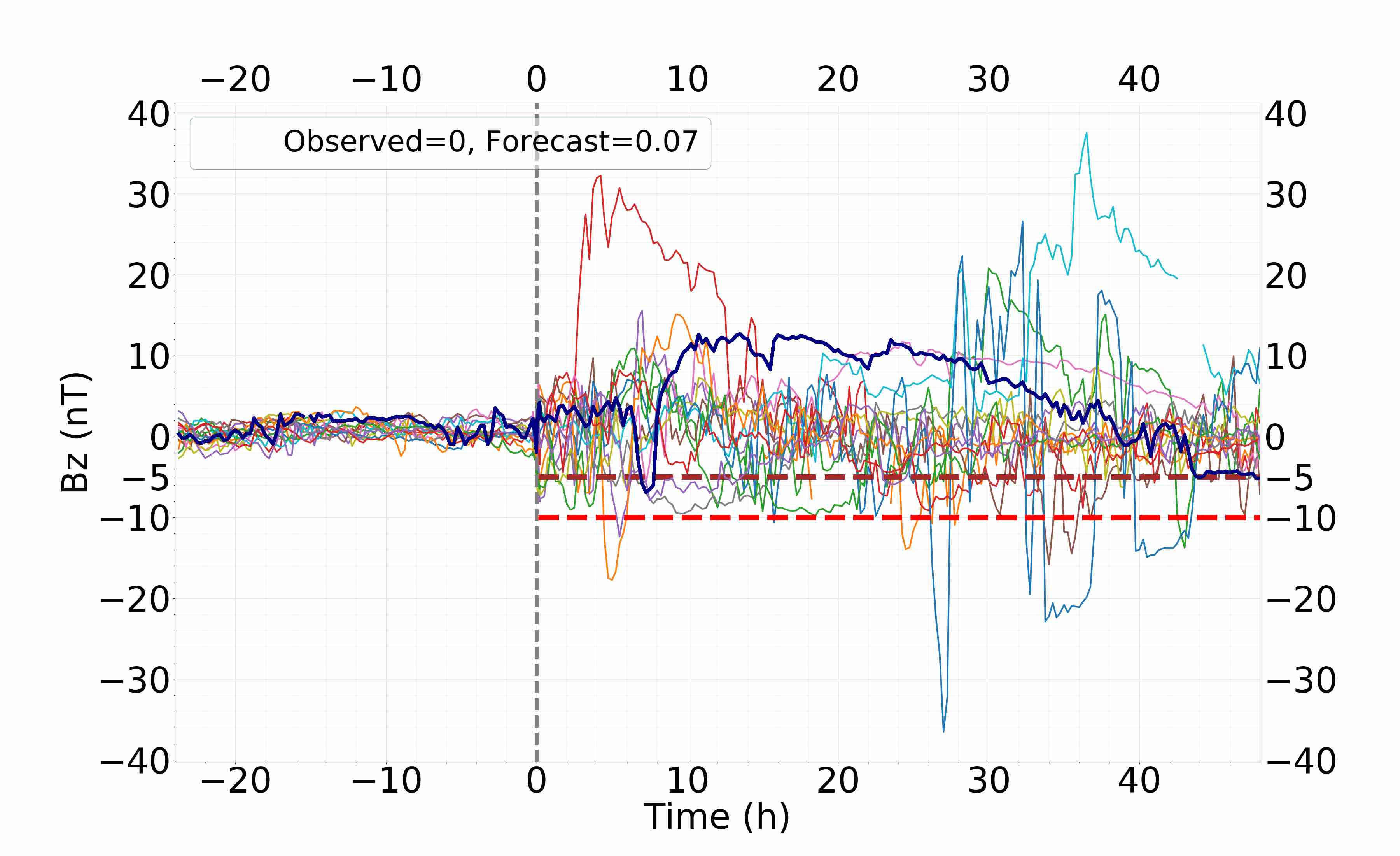}
    \caption{Probabilistic forecast for the event following the 29 March 2011 shock shown in Figure~\ref{fig:reference event-II} for TWWCP (1,0) and weights (0, 1, 0, 0).}
     \label{fig:forecast1a}
  \end{figure*}

\begin{figure*}[htbp!]
  \centering
    \includegraphics[width=1.0\linewidth]{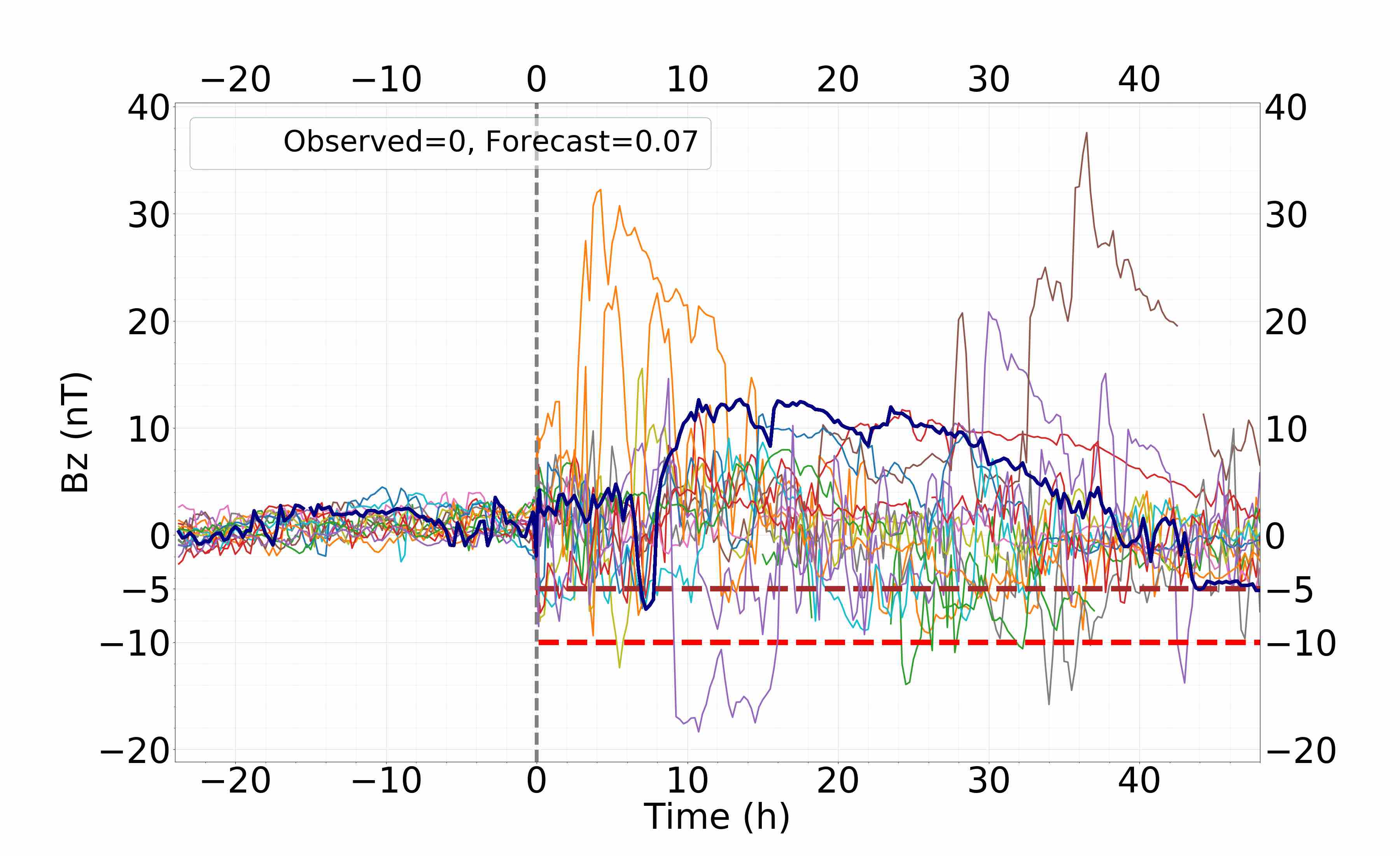}
    \caption{Same as Figure~\ref{fig:forecast1a} but for weights (1 , 0.25, 0, 0).}
     \label{fig:forecast2a}
  \end{figure*}

For the baseline set of weights (0, 1, 0, 0) and TWWCP (1,0), a probabilistic forecast of 0.07 (Figure~\ref{fig:forecast1a}) is obtained. This probabilistic forecast is almost the ideal forecast. However, recognizing the pattern of optimum probabilistic forecast is the primary goal. The trend observed from the first reference event is that adding more parameters and putting more weight into the variations closer to the shock generally improves the probabilistic forecast. However, the probabilistic forecast worsens when the variations of V$_{x}$ in the training window is considered. These assumptions developed from the analysis of the first reference event are put to test here. For the set of weights (1, 0.25, 0, 0) (Figure~\ref{fig:forecast2a}) with TWWCP (1,0) a probabilistic forecast of 0.07 is obtained. The probabilistic forecast is not improved by adding a new parameter (B) like the first reference event but is not worsened either.

\begin{figure*}[htbp!]
\centering
    \includegraphics[width=1.0\linewidth]{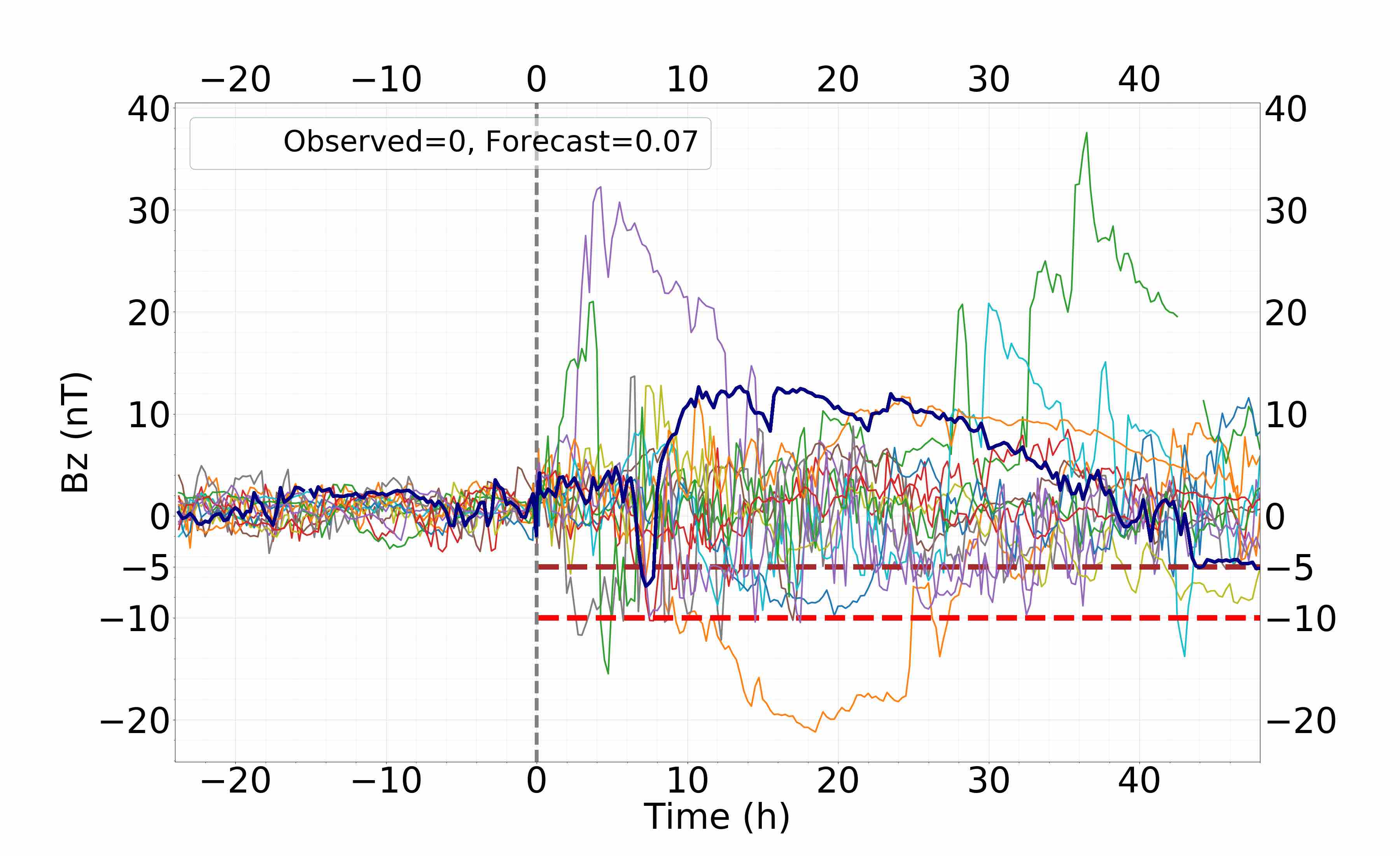}
    \caption{Same as Figure~\ref{fig:forecast1a} but for TWWCP (\( \frac{1}{4}\ ,\frac{3}{4}\)) and weights (1, 1, 0.5, 0).}
     \label{fig:forecast5a}
\end{figure*}

Next, the third parameter N$_{p}$  is added and more weight is put into the variations closer to the shock. For the set of weights (1, 1, 0.5, 0) with TWWCP (\( \frac{1}{3}\ ,\frac{2}{3}\)) and (1, 1, 0.5, 0) with TWWCP (\( \frac{1}{4}\ ,\frac{3}{4}\)) (Figure~\ref{fig:forecast5a}), probabilistic forecasts of 0.07 is obtained for both the cases.

However, for the set of weights (1, 1, 1, 1) with TWWCP (\( \frac{1}{2}\ ,\frac{1}{2}\)) probabilistic forecast of 0.20 is obtained. Similarly for the set of weights including V$_{x}$ like (1, 0.5, 0.25, 0.25), (1, 1, 0, 0.5), and (0, 1, 0.25, 0.25) with fixed TWWCP (\( \frac{1}{2}\ ,\frac{1}{2}\)), probabilistic forecasts of 0.20, 0.13, and 0.13 are obtained respectively which confirms the assumption that adding the parameter V$_{x}$  generally makes the forecast worse. Now, for the same selected threshold of 0.40 as the first reference event, all of the predictions discussed above for the second reference event are considered a NO and the outcome is the correct prediction.

The pattern observed through the analysis of reference events (only 2 shown in this section) holds true for both positive (B$_{z}$\textless $-10$~nT periods of 3 consecutive hours or more in forecast window) and negative (no B$_{z}$\textless $-10$~nT period of 3 consecutive hours or more in forecast window) events. The probable best combination corresponds to maximum weights assigned to parameters B and B$_{z}$, inclusion of parameter N$_{p}$ with reduced weight in comparison with B and B$_{z}$, exclusion of parameter V$_{x}$ in identifying the closest matches, and  increased weight in the 30 minutes interval around the arrival of fast-forward shocks.

\section{Results} \label{sec:results}

Based on the construction of our probabilistic forecast model and selection of sets of weights and TWWCPs, several thousands of different combinations can be used to build the probabilistic forecast model. However, the main goal of this study is to show that such a probabilistic and threshold-based forecast model has skills and lay the groundwork for more detailed experimentation in the future.

\subsection{Best Probabilistic Forecast for All 100 Reference Events}
 
We try out different combinations of sets of weights and TWWCPs for the first set of 24 reference events in a way similar to what is shown for the two events in Section~\ref{sec:example}. We then determine the probabilistic forecast for the forecast parameter (B$_{z}$\textless $-10$~nT for a period of 3 consecutive hours or more). The error of each probabilistic forecast for each reference event is calculated, as well as the RMS errors. This allows us to select the seven best performing combinations as well as the baseline forecast based only on B$_{z}$ and no shock info (Table~\ref{tab:1}).

 \begin{table}[htbp!]
  \centering
  \caption{Seven best performing and the baseline combinations of sets of weights and TWWCPs for the first set of reference events and their corresponding RMS errors of probabilistic forecasts.}
    \begin{tabular}{cccccc}
     \hline
      \multicolumn{4}{c}{Set of Weights}&     & \\\hline
    B   & B$_{z}$      & N$_{p}$    & V$_{x}$     & TWWCP & RMSE (Set-I) \\\hline
    0     & 1     & 0     & 0     & 1,0   & 0.387 \\\hline
    1  & 0.5     & 0     & 0     & 1/4,3/4 & 0.318 \\\hline
    1   & 0.5     & 0.25  & 0     & 1/4,3/4 & 0.310 \\\hline
    1   & 0.5     & 0.5   & 0     & 1/3,2/3 & 0.311 \\\hline
    1   & 0.5     & 0.5   & 0     & 1/4,3/4 & 0.309 \\\hline
    1     & 1     & 0     & 0     & 1/3,2/3 & 0.303 \\\hline
    1     & 1     & 0.25  & 0     & 1/4,3/4 & 0.289 \\\hline
    1     & 1     & 0.5   & 0     & 1/4,3/4 & 0.310 \\\hline
    \end{tabular}%
  \label{tab:1}%
\end{table}%

Looking at this list of 8 combinations, it is observed that the highest RMSE of 0.387 belongs to the baseline set of weights (0, 1, 0, 0) with TWWCP (1,0) and the lowest RMSE of 0.289 belongs to the set of weights (1, 1, 0.25, 0) with TWWCP (\( \frac{1}{4}\ ,\frac{3}{4}\)). An important thing to note is that RMS errors for these eight combinations are well below 0.5 which means each of them is able to provide a probabilistic forecast substantially better than a random forecast. The best non-skilled forecast can be obtained by assigning a probability of 0.23 to all fast-forward shock events resulting in a RMSE of 0.421. The baseline combination has a RMSE which is 8.08{\%} lower than this, whereas the best RMSE is significantly lower (by 31{\%}).

Then, for each of these eight combinations, we calculate the errors of forecasts (Table~\ref{tab:2}) for the other three sets of reference events (25, 25, and 26 reference events respectively).

\begin{table}[htbp!]
  \centering
  \caption{Seven best performing and the baseline combinations of sets of weights and TWWCPs for the other three sets of reference events and their corresponding RMS errors of probabilistic forecasts.}
    \begin{tabular}{cccccccc}
     \hline
     \multicolumn{4}{c}{Set of Weights}&     & \multicolumn{3}{c}{RMSE} \\\hline
    B  &B$_{z}$       &N$_{p}$       &V$_{x}$       &TWWCP       & Set-II & Set-III & Set-IV \\\hline
    1   & 0.5     & 0     & 0     & 1/4,3/4 & 0.3161 & 0.3773 & 0.4716 \\\hline
    1   & 0.5     & 0.25  & 0     & 1/4,3/4 & 0.3331 & 0.3795 & 0.4420 \\\hline
    1   & 0.5     & 0.5   & 0     & 1/3,2/3 & 0.3564 & 0.3564 & 0.4369 \\\hline
    1   & 0.5     & 0.5   & 0     & 1/4,3/4 & 0.3403 & 0.3612 & 0.4287 \\\hline
    0     & 1      & 0     & 0     & 1,0   & 0.3814 & 0.3268 & 0.4427 \\\hline
    1     & 1     & 0     & 0     & 1/3,2/3 & 0.3128 & 0.3588 & 0.4618 \\\hline
    1     & 1     & 0.25  & 0     & 1/4,3/4 & 0.3435 & 0.3838 & 0.4551 \\\hline
    1     & 1     & 0.5   & 0     & 1/4,3/4 & 0.3356 & 0.3915 & 0.4629 \\\hline
    \end{tabular}%
  \label{tab:2}%
\end{table}%

Finally, we calculate the average RMSE for each of these 8 combinations for all the 100 reference events (Table~\ref{tab:3}). Then we pick out 5 combinations from these 8 combinations, 4 in terms of lowest average RMS errors and the fifth one is the baseline combination. The lowest average RMSE belongs to the set of weights (1, 1, 0, 0) with TWWCP (\( \frac{1}{3}\ ,\frac{2}{3}\)) which has an average RMSE of 0.3592.

\begin{table}[htbp!]
  \centering
  \caption{Average RMS errors of probabilistic forecasts corresponding to the seven best performing and the baseline combinations of sets of weights and TWWCPs for the four sets of reference events.}
    \begin{tabular}{cccccc}
     \hline
    \multicolumn{4}{c}{Set of Weights}&    &\\\hline
    B   & B$_{z}$      & N$_{p}$     & V$_{x}$     & TWWCP & Average RMSE \\\hline
    1   & 0.5     & 0     & 0     & 1/4,3/4 & 0.3708 \\\hline
    1   & 0.5     & 0.25  & 0     & 1/4,3/4 & 0.3662 \\\hline
    1   & 0.5     & 0.5   & 0     & 1/3,2/3 & 0.3652 \\\hline
    1   & 0.5     & 0.5   & 0     & 1/4,3/4 & 0.3597 \\\hline
    0     & 1      & 0     & 0     & 1,0   & 0.3845 \\\hline
    1     & 1     & 0     & 0     & 1/3,2/3 & 0.3592 \\\hline
    1     & 1     & 0.25  & 0     & 1/4,3/4 & 0.3678 \\\hline
    1     & 1     & 0.5   & 0     & 1/4,3/4 & 0.3750 \\\hline
    \end{tabular}%
  \label{tab:3}%
\end{table}%

\subsection{Threshold-Based Probabilistic Forecast}

Then we move on to the second step of the probabilistic model. For each of these 5 combinations, we try to find the best threshold criterion out of 15 (see Section~\ref{ssec:threshold}) corresponding to the optimum Heidke Skill Score (HSS). To find the optimum HSS, we consider the probabilistic forecasts of the 100 reference events. Looking at the scores (Table~\ref{tab:4}), we see that the lowest best HSS of 0.31 belongs to the baseline combination for a threshold of 0.33 or \( \frac{5}{15} \). The highest best HSS of 0.44 belongs to the set of weights (1, 1, 0, 0) with TWWCP (\( \frac{1}{3}\ ,\frac{2}{3}\)) for a threshold of 0.40 or \( \frac{6}{15} \). This HSS represents a 42{\%} improvement over the baseline HSS.

\begin{table}[htbp!]
  \centering
  \caption{Threshold criterion corresponding to the optimum Heidke Skill Score for the four best performing and the baseline sets of weights and TWWCPs.}
    \begin{tabular}{ccccccc}
     \hline
    \multicolumn{4}{c}{Set of Weights} &       & \multicolumn{2}{c}{B$_{z}$\textless -10~nT} \\\hline
    B    & B$_{z}$     & N$_{p}$    & V$_{x}$    & TWWCP & Best HSS & Threshold Criterion \\\hline
    0     & 1      & 0     & 0     & 1,0   & 0.31  & 0.33 \\\hline
    1     & 1     & 0     & 0     & 1/3,2/3 & 0.44  & 0.40 \\\hline
    1   & 0.5     & 0.25  & 0     & 1/4,3/4 & 0.42  & 0.40 \\\hline
    1   & 0.5     & 0.5   & 0     & 1/4,3/4 & 0.39  & 0.33 \\\hline
    1   & 0.5     & 0.5   & 0     & 1/3,2/3 & 0.34  & 0.33 \\\hline
    \end{tabular}%
  \label{tab:4}%
\end{table}%

Now, we use this threshold of 0.40 and the set of weights (1, 1, 0, 0) with TWWCP (\( \frac{1}{3}\ ,\frac{2}{3}\)) to find the HSS for 10 randomly chosen sets, each with 166 reference events (166 events account for slightly more than \( \frac{1}{3} \) of our database (Table~\ref{tab:5}). In this phase, only one forecast is generated at a time by removing only the event in consideration from the database.

\begin{table}[htbp!]
  \centering
  \caption{Average contingency table and HSS for 10 randomly chosen sets of 166 reference events each.}
    \begin{tabular}{ccccc}
      \hline
          & Forecast & \multicolumn{2}{c}{Observed} &  \\\hline
    Average set &       & YES   & NO    & Marginal Total \\\hline
          & YES   & 18 (\textit{Hit})     & 4 (\textit{False Alarm})     & 22 \\\hline
          & NO    & 12 (\textit{Miss})     & 132 (\textit{Correct Negative})    & 144 \\\hline
          & Marginal Total & 30 (\textit{Observed YES})     & 136 (\textit{Observed NO})    & 166 (\textit{Total}) \\\hline
          & HSS & \multicolumn{2}{c}{0.64}       &\\\hline
    \end{tabular}%
  \label{tab:5}%
\end{table}%

If we examine these 10 sets, each with 166 threshold-based forecasts (Table~\ref{tab:5}), the best combination provides on average 22 YES forecasts, 18 of them are correct predictions and 4 are false alarms (average False Alarm Ratio of 0.18 indicating that on average in \( \frac{2}{11} \) of the forecast for B$_{z}$ periods, strong and long-duration B$_{z}$ were not observed). Average Threat Score or Critical Success Index is 0.53 indicating that on average slightly greater than \( \frac{1}{2} \) of strong and long-duration B$_{z}$ periods (observed and/or predicted) were correctly forecast. However, the model correctly predicts on average 18 out of 30 events that occurred (average Probability of Detection Yes is 0.60 indicating that on average 60{\%} of the observed B$_{z}$ periods were correctly predicted). The average Bias (Frequency) score is 0.73 indicating strong under-forecasting of strong and long-duration B$_{z}$ periods. It is evident as the model provides on average 144 NO forecasts compared to 22 YES forecasts.

\section{Discussion, Conclusions and Future Work} \label{sec:conclusion}

We developed a two step threshold-based probabilistic model for forecasting B$_{z}$\textless -10~nT periods of 3 consecutive hours or more in the 48 hours after a fast-forward shock arrival at L1. In the first step, the model provides a probabilistic forecast based on an analogue ensemble approach. A pre-defined threshold is imposed on the forecast in the second step. Any forecast exceeding this threshold is considered a YES and any forecast below this threshold is considered a NO. The forecast capability of the model is evaluated using a skill score approach.

To construct the model, we use the association between fast-forward shocks and strong and long-duration southward B$_{z}$ periods as the basis. We select the interval of 24 hours prior to the shock to 0.25 hours after the shock arrival as the training window and the interval of 0.25 hours to 48 hours after the shock arrival as the forecast window. For any reference event, we use the training window to determine the closest matches to the solar wind conditions to be forecasted and use the observations of the post-shock intervals of these closest matches to make a probabilistic forecast for the reference event. These closest matches are found through quantifying variations of four solar wind and interplanetary magnetic field parameters (B, B$_{z}$, N$_{p}$, V$_{x}$) in the training window using the RMSE approach. We split the training window into two intervals (24 hours to 0.25 hours prior to the shock and 0.25 hours prior to the shock to 0.25 hours after the shock arrival). For the former interval, we bin the magnetic field and plasma data into 15-minute averages. RMS errors of the four parameters in both intervals are multiplied with weighting constants. We also assign different weights to the four parameters. We try out different combinations of sets of weights and training window weighting constant pairs for a set of reference events and determine the RMSE of forecast for each combination. Analyzing these errors for 100 reference events, we select 5 combinations (4 in terms of lowest average RMS errors and 1 as baseline). Finally, for each of these 5 combinations the threshold criterion corresponding to the optimum Heidke Skill Score (HSS) is identified. The best HSS of 0.44 for a threshold criterion of 0.40 belongs to the set of weights (1, 1, 0, 0) with TWWCP (\( \frac{1}{3}\ ,\frac{2}{3}\)).

Examining the forecasts provided by this combination for 1660 random events, we see an average Critical Success Index of 0.53 indicating that on average slightly greater than \( \frac{1}{2} \) of strong and long-duration B$_{z}$ periods (observed and/or predicted) were correctly forecast. 

\citet{Riley:2017} outlined a pattern recognition technique for forecasting solar wind parameters. Forecast of any solar wind parameter into the future is made based on an ensemble of prior observations and extrapolation of information provided by these observations to make a probabilistic prediction of current time. Our baseline combination of weights (0, 1, 0, 0) with TWWCP (1,0) is closely related to the method adopted by them to forecast B$_{z}$. This particular combination shows capability for predicting B$_{z}$ periods subject to certain criteria. In our limited study, this has proven to be one of the best methods for forecasting B$_{z}$\textless -5~nT periods of 1h. 

However, this method does not perform well for long-duration and intense southward B$_{z}$ periods, as was the focus here. 
The modifications applied to our threshold-based probabilistic model picks out the combination (1, 1, 0, 0) with TWWCP (\( \frac{1}{3}\ ,\frac{2}{3}\)) and the threshold criterion 0.40 as the best method for forecasting B$_{z}$\textless -10~nT periods of 3 consecutive hours or more in the 48 hours after a fast-forward shock arrival at L1. The average RMSE associated with this combination is 0.3592 (for 100 reference events), which is 14.7{\%} lower than the RMSE of 0.421 of the non-skilled forecast. The non-skilled forecast refers to assigning a 0.23 probability to any fast-forward shock being followed by an intense storm within 48 hours after its arrival. Quantitatively this combination is 28{\%} better compared to the coin-flipping/random model (RMSE of 0.50). 
              
The model would provide on average a 14-hour warning of an upcoming intense southward B$_{z}$ period. This puts it in-between the L1/nowcast forecast ($\sim$30 minutes) and those based on solar/coronal data (1-3 days). Preliminary statistics shows that the model has significant skills. The basis of the model is the association between fast-forward shocks and strong and long-duration southward B$_{z}$ periods. Our study validates this close association as we find 76{\%} of strong and long-duration southward B$_{z}$ periods over a 22.4-year span to be preceded by fast-forward shocks. Thereby at best we would expect the model to successfully predict 76{\%} southward B$_{z}$ periods. On a similar note, we can make the database more robust by including data from STEREO and ACE spacecraft (especially as SC24 is different and weaker from SC23, \citep[]{McComas:2013}) and also shock-like discontinuities. Another foreseeable improvement would be to try out more combinations of sets of weights and training window weighting constant pairs to further minimize the error of forecasts. We only tried a handful of combinations based on the pattern of improving probabilistic forecasts developed through analyzing multiple reference events. The final logical improvement would be to develop a dynamic forecasting model. Our analysis shows that including post-shock information and putting more weight into solar wind variations closer to the shock have the tendency to significantly improve the probabilistic forecasts. Thereby this proposed dynamic model will give a probabilistic forecast 0.25 hours after a fast-forward shock arrival at L1. Then this forecast will be constantly updated up to a certain period through monitoring post-shock solar wind conditions. However, this evolving forecasting period can be a challenge to determine as our findings suggest that 10{\%} of strong and long-duration southward B$_{z}$ periods occur within 1.85 hours after the shock arrival. Thereby a proper balance needs to be established between advance warning and accuracy of forecasts.                  

In closing, we have developed a threshold-based probabilistic forecast model using a ``superposed epoch analysis''-like approach. The limited trials to identify the best method to predict prolonged and intense southward B$_{z}$ periods have shown encouraging signs. Preliminary statistics shows that the forecasts provided by the model are measurably better than a random model. However, the model has not been entirely successful in providing accurate and actionable forecasts. It is also uncertain how much improvement would be achieved through adopting the aforementioned refinements. We are planning to conduct a more extensive study in the future to evaluate and develop the model further. The ultimate goal is to devise a model that can function as a real-time forecasting tool. Given the underlying complexity of the solar wind and a forecast parameter like B$_{z}$ which has limited inherent predictability, this model is definitely a step in the right direction.   

\begin{acknowledgments}
The authors acknowledge use of NASA/GSFC's Space Physics Data Facility's OMNIWeb (or CDAWeb or ftp) service, OMNI data available at https://omniweb.gsfc.nasa.gov/, \textit{Wind} data available at https://cdaweb.sci.gsfc.nasa.gov/index.html/, the Heliospheric shock database, generated and maintained at the University of Helsinki which can be found at: http://ipshocks.fi/, and shock database of Harvard Smithsonian Center for Astrophysics which can be found at: https://www.cfa.harvard.edu/shocks/ and sincerely thank the corresponding teams for their open data policy. The research for this manuscript was supported by the following grants: NSF AGS-1435785, AGS-1433086 and AGS-1433213 and NASA NNX15AB87G, NNX15AU01G and NNX16AO04G. R.M.W. acknowledges support from NASA grant NNX15AW31G and NSF grant AGS1622352.
\end{acknowledgments}


\end{document}